\documentclass[twocolumn]{aastex63}

\usepackage{siunitx}
\usepackage{subfigure}
\usepackage{pgfplots}
\pgfplotsset{compat=newest}

\submitjournal{ApJS}
\shorttitle{\textsc{GRETOBAPE} gas-phase reaction network}
\shortauthors{Tinacci et al.}
\usepackage{comment}
\usepackage{hyperref}
\usepackage[version=4]{mhchem}
\usepackage{chemfig}
\usepackage{multirow}

\begin{document}
\title{The \textsc{GRETOBAPE} gas-phase reaction network: the importance of being exothermic}
\correspondingauthor{Lorenzo Tinacci}
\email{Lorenzo.Tinacci@univ-grenoble-alpes.fr}
\author[0000-0001-9909-9570]{Lorenzo Tinacci}
\affiliation{Université Grenoble Alpes, CNRS, IPAG, 38000 Grenoble, France}
\affiliation{Dipartimento di Chimica and Nanostructured Interfaces and Surfaces (NIS) Centre,\\ Università degli Studi di Torino, via P. Giuria 7, 10125 Torino, Italy}
\author[0000-0002-4874-838X]{Simón Ferrada-Chamorro}
\affiliation{Université Grenoble Alpes, CNRS, IPAG, 38000 Grenoble, France}
\author[0000-0001-9664-6292]{Cecilia Ceccarelli}
\affiliation{Université Grenoble Alpes, CNRS, IPAG, 38000 Grenoble, France}
\author[0000-0002-2457-1065]{Stefano Pantaleone}
\affiliation{Dipartimento di Chimica and Nanostructured Interfaces and Surfaces (NIS) Centre,\\ Università degli Studi di Torino, via P. Giuria 7, 10125 Torino, Italy}
\affiliation{Dipartimento di Chimica, Biologia e Biotecnologie, Università di Perugia, 06123 Perugia, Italy}
\author[0000-0001-5393-9554]{Daniela Ascenzi}
\affiliation{Dipartimento di Fisica, Università di Trento, 38123 Povo, Italy}
\author[0000-0002-5524-8068]{Andrea Maranzana}
\affiliation{Dipartimento di Chimica and Nanostructured Interfaces and Surfaces (NIS) Centre,\\ Università degli Studi di Torino, via P. Giuria 7, 10125 Torino, Italy}
\author[0000-0001-5121-5683]{Nadia Balucani}
\affiliation{Université Grenoble Alpes, CNRS, IPAG, 38000 Grenoble, France}
\affiliation{Dipartimento di Chimica, Biologia e Biotecnologie, Università di Perugia, 06123 Perugia, Italy}
\affiliation{Osservatorio Astrofisico di Arcetri, Largo E. Fermi 5, 50125 Firenze, Italy}
\author[0000-0001-8886-9832]{Piero Ugliengo}
\affiliation{Dipartimento di Chimica and Nanostructured Interfaces and Surfaces (NIS) Centre,\\ Università degli Studi di Torino, via P. Giuria 7, 10125 Torino, Italy}

\begin{abstract}
The gas-phase reaction networks are the backbone of astrochemical models. 
However, due to their complexity and non-linear impact on the astrochemical modeling, they can be the first source of error in the simulations if incorrect reactions are present. 
Over time, following the increasing number of species detected, astrochemists have added new reactions, based on laboratory experiments and quantum mechanics (QM) computations as well as reactions inferred by chemical intuition and similarity principle.
However, sometimes no verification of their feasibility in the interstellar conditions, namely their exothermicity, was performed. 
In this work, we present a new gas-phase reaction network, \textsc{GRETOBAPE}, based on the KIDA2014 network and updated with several reactions, cleaned from endothermic reactions not explicitly recognized as such.
To this end, we characterized all the species in the \textsc{GRETOBAPE} network with accurate QM calculations. 
We found that $\sim 5$\% of the reactions in the original network are endothermic although most of them are reported as barrierless. 
The reaction network of Si-bearing species is the most impacted by the endothermicity cleaning process.
We also produced a cleaned reduced network, \textsc{GRETOBAPE-red}, to be used to simulate astrochemical situations where only C-, O-, N- and S- bearing species with less than 6 atoms are needed.
Finally, the new \textsc{GRETOBAPE} network, its reduced version, as well as the database with all the molecular properties are made publicly available.
The species properties database can be used in the future to test the feasibility of possibly new reactions. 

\end{abstract}


\section{Introduction } \label{sec:intro}

Since the discovery of the first molecules in the interstellar medium (ISM) in the visible spectrum \citep{swings1937considerations,dunham1937interstellar,mckellar1940evidence}, the astrochemical community has been trying to understand their origin and how they can survive the harsh interstellar conditions. 
With the evolution of astronomical facilities and, more specifically, with the advent of radio telescopes, new and more complex molecules have been detected in various astronomical objects with an almost constant discovery rate of $\sim4-6$ new molecules per year \citep{McGuireCensus2021}. 
To date, more than 270 molecules, composed of 19 different elements, have been detected in the interstellar and circumstellar medium, and this number is steadily increasing. 
The list of detected interstellar species can be found in the Cologne Database for Molecular Spectroscopy (CDMS) \citep{endres2016cologne}\footnote{\url{https://cdms.astro.uni-koeln.de/classic/molecules}}, in the \cite{McGuireCensus2021} census (and the related Python package\footnote{\url{https://github.com/bmcguir2/astromol}}), and in The Astrochymist website\footnote{\url{http://www.astrochymist.org/astrochymist_ism.html}}.

Understanding how interstellar molecules are formed, destroyed and connected, in addition to be interesting per se, can help inferring the properties, history, and evolution of the astronomical objects in which they have been detected. 
Thus, in parallel with the identification of molecules and the measurements of their abundances in different astronomical objects, soon a community dedicated to developing astrochemical models began to play an important role \citep{herbst1973formation,prasad1980model}. 
A crucial element of astrochemical models is the gas-phase reaction network, namely the lists of the gas-phase chemical reactions with their rate constants, that describe how efficient the reactions are as a function of the temperature, and product branching ratios.

The increasing number of detected species has prompted modelers and chemists to add new reactions to the existing networks in order to reproduce the observed species and their abundances.
In this way, the number of reactions, reactants and products have been increasing each time a new molecule was detected, which also led to an increase in the network size and complexity. 
Nowadays, the publicly available astrochemical gas-phase networks, KIDA\footnote{\url{http://kida.obs.u-bordeaux1.fr/}} \citep{wakelam2014_Kida_uva} and UMIST\footnote{\url{http://udfa.ajmarkwick.net/}} \citep{mcelroy2013umist}, contains about 8000 reactions involving more than 500 species.

Unfortunately, the vast majority ($\geq 80$\%) of the rates and branching ratios of the reported gas-phase reactions has not been measured or computed and is often based on approximate estimates \citep[\textit{e.g.} using the Capture Theory:][]{su1982parametrization,Herbst2006,woon2009quantum,loison2013gas}, educated guesses on similarity principles or simple chemical intuition.
In addition, even when some experimental measurements are available, the estimated rate constants may have substantial uncertainties as they are often based on experiments at room temperature \citep[\textit{e.g.}][]{anicich2003index}. 
The uncertainty on the reaction rates is a well-known problem in the astrochemistry community. 
In parallel with the efforts of studying reactions via experiments and theoretical computations, sensitivity analysis studies have provided the impact of these uncertainties on the abundances predicted by astrochemical models \citep[\textit{e.g.}][]{wakelam2006chemical, penteado2017}.

One possible and mandatory first step to improve the reliability of the gas-phase reaction networks, that has never been performed systematically on the astrochemical networks is to estimate the exo/endothermicity of the gas-phase reactions, using accurate physico-chemical quantum mechanical (QM) data for each species.
Indeed, given the ISM conditions (temperatures usually less than $\sim$100-200 K), strongly endothermic reactions are inhibited and, therefore, can be excluded from the astrochemical networks without complex and time-consuming reaction transition states studies or experiments \citep{smith2006reactions,smith2011laboratory}. 
This verification is now possible, thanks to the availability of two theoretical studies that together provide reliable QM data (electronic state, electron spin multiplicity, geometry, harmonic frequencies, absolute electronic energy, and dipole moment) for the totality of the species in the KIDA network.
The first one by \cite{woon2009quantum} computed the QM data for a large number of neutral molecules involved in the KIDA reaction network.
The second study by \cite{ISM_cations} complemented the first one and provided QM data for all the ions and the remaining neutral species of the KIDA network.
Please note that the two studies have been carried out at the same electronic level of theory, so that they can be reliably used for computing the exo/endothermicity of all the reactions.
Also, \cite{ISM_cations} created a database\footnote{\url{https://aco-itn.oapd.inaf.it/aco-public-datasets/theoretical-chemistry-calculations}} with the above QM data for all the cations.

The KIDA and UMIST networks contain such a large number of reactions because their goal is to reproduce the observed abundances of large molecules, which have usually low (between $10^{-8}$ and $10^{-12}$) abundances.
However, in some cases, such as astrochemical models coupled with hydrodynamical simulations of cloud collapse or protoplanetary disks formation and evolution, the goal is to reliably reproduce only the most abundant species.
In these cases, the use of large reactions networks is not only useless but also detrimental, because it demands spareable computing time.
Indeed, several authors have proposed reaction network reduction techniques in the last few years, each using different criteria and focusing on specific goals \citep{oppenheimer1974fractional,nelson1999stability,glover2010modelling,grassi2013chemical,grassi2021reducing}. 

\begin{deluxetable*}{llccc}[!ht]
\tablecaption{ZPE-corrected reaction energy difference, \textit{i.e.}, $\Delta (\Delta \mathrm{H}(0))$, between previous literature and our data.}
\label{tab:reactions_bench}
\tablehead{\multicolumn{2}{c}{Reaction} & $\Delta (\Delta \mathrm{H}(0))$ & \multirow{2}{*}{Their Level of Theory} & \multirow{2}{*}{Reference} \\ Reactants & Products  & [kJ/mol] &  & }
\startdata
C$_2$H$_5$ + O & C$_2$H$_4$ + OH  &  1.26 & \multirow{3}{*}{CCSD(T)/aug-cc-pVTZ//B2PYLYP/aug-cc-pVTZ} & \multirow{3}{*}{\cite{Vazart_Cecarrelli_2020_acetaldehyde}} \\
C$_2$H$_5$ + O & CH$_3$CHO + H    &  1.67 & &  \\
C$_2$H$_5$ + O & H$_2$CO + CH$_3$ &  1.23 & &  \\
\hline
CH + CH$_3$OH  & C$_2$H$_4$ + OH  &  0.36 & \multirow{3}{*}{CCSD(T)/aug-cc-pVTZ//B2PYLYP/aug-cc-pVTZ} & \multirow{3}{*}{\cite{Vazart_Cecarrelli_2020_acetaldehyde}} \\
CH + CH$_3$OH  & CH$_3$CHO + H    &  0.77 & &  \\
CH + CH$_3$OH  & H$_2$CO + CH$_3$ &  0.34 & &  \\
\hline
SiH + S        & SiS + H          &  4.29 & \multirow{2}{*}{CCSD(T)/aug-cc-pVTZ//B3LYP/aug-cc-pVTZ} & \multirow{2}{*}{\cite{Rosi_2018_SiS}} \\
SiH + S$_2$    & SiS + HS         & -1.85 & &  \\
\hline
Si + SH        & SiH + S          &  -3.92 & MRCI/CBS & \cite{mota2021sis}
\enddata
\tablecomments{
The minor differences with respect to \cite{Vazart_Cecarrelli_2020_acetaldehyde} and \cite{Rosi_2018_SiS} can be ascribed to the different optimization levels and also the spin contamination for all open-shell species, which was corrected in our case by using the Restricted-Open formalism.}
\end{deluxetable*}

The goal of the present article is twofold.
The \textit{first goal} is to produce a gas-phase reaction network "cleaned" from endothermic reactions not reported as such.
To this purpose, we perform a systematic study of the endo/exo thermicity of all reactions present in a new network, \textsc{GRETOBAPE}, which is based on the KIDA2014 one and updated through the years by our group.
In the process, we  also verify that each species present in the network is not a sink and that the obvious reactions with the most abundant molecular cations are present.
The \textit{second goal} is to produce a reduced network from the cleaned one, which only includes H-, C-, N-, O-, and S-bearing species with not more than six atoms, to be employed when the use of a complete network is unnecessary.

The article is organized as follows. 
Section \ref{sec:methodology} details the adopted tools to deal with the reaction network, the cleaning procedure, and the computational methodology used to characterize all the chemical species present in the network. 
In Section \ref{sec:results}, we describe the new reaction networks, \textsc{GRETOBAPE} and \textsc{GRETOBAPE-red}, respectively resulting from the cleaning and reduction processes, as well as a list of the obvious reactions possibly missing in the network. 
In Section \ref{sec:implications}, we discuss the obtained results and their impact on astrochemical models of cold molecular clouds and warm molecular outflow shocks simulations. 
Finally, in Section \ref{sec:conclusions}, we present our conclusions, and in Section \ref{sec:online-db} we provide the hyperlinks to the publicly available online databases and reaction networks produced in this work.

\section{Methodology} \label{sec:methodology}


\subsection{Computational details for QM calculation}\label{subsec:meth-QM-calc}

All the species present in the astrochemical reaction network were characterized via QM calculations to obtain their optimized structure, absolute electronic energy, and Zero Point Energy (ZPE) correction. 
Most of the neutral species structures and their corresponding electronic spin multiplicity ($m_s$) in the ground state, a fundamental piece of information for the electronic QM calculations, are taken from \cite{woon2009quantum} while for the cations, the information is taken from our previous work \citep{ISM_cations}. 
For the chemical species that have not been theoretically characterized yet, we inferred here the structure and $m_s$ using the same approach introduced in \cite{ISM_cations}.
The list of those species with their chemical data are reported in the Supporting Material, in the file \texttt{species.zip}.

In order to have all species characterized at the same level of theory, \textit{i.e.}, so to have a consistent method, we used the DFT M06-2X \citep{M06-2X} coupled with the triple-$\zeta$ Dunning's correlation consistent basis set (cc-pVTZ) \citep{kendall1992dunning,woon1993gaussian} for geometry optimization and harmonic frequency calculations. 
The electronic absolute energy was refined at a CCSD(T) \citep[][for closed shell species]{knowles1993coupled} or RO-CCSD(T) \citep[][for open shell species]{watts1993coupled} level of theory, in conjunction with an augmented triple-$\zeta$ correlation consistent basis set \citep[aug-cc-pVTZ;][]{kendall1992dunning}. 
All the molecular structures and properties computed with the adopted methodology were benchmarked in our previous work \citep{ISM_cations}. 
All calculations were carried out with the Gaussian16 program \citep{g16} and we kept the default values set up in the program. 

\tikzset{
square matrix/.style={
    matrix of math nodes,
    column sep=-\pgflinewidth, 
    row sep=-\pgflinewidth,
    nodes in empty cells,
    nodes={draw,minimum width=#1,minimum height=.5*#1,anchor=center,inner sep=0pt,align=center},
    },
square matrix/.default=0.8cm
}

\begin{figure*}[!ht]
\begin{center}
\includegraphics[width=0.99\textwidth]{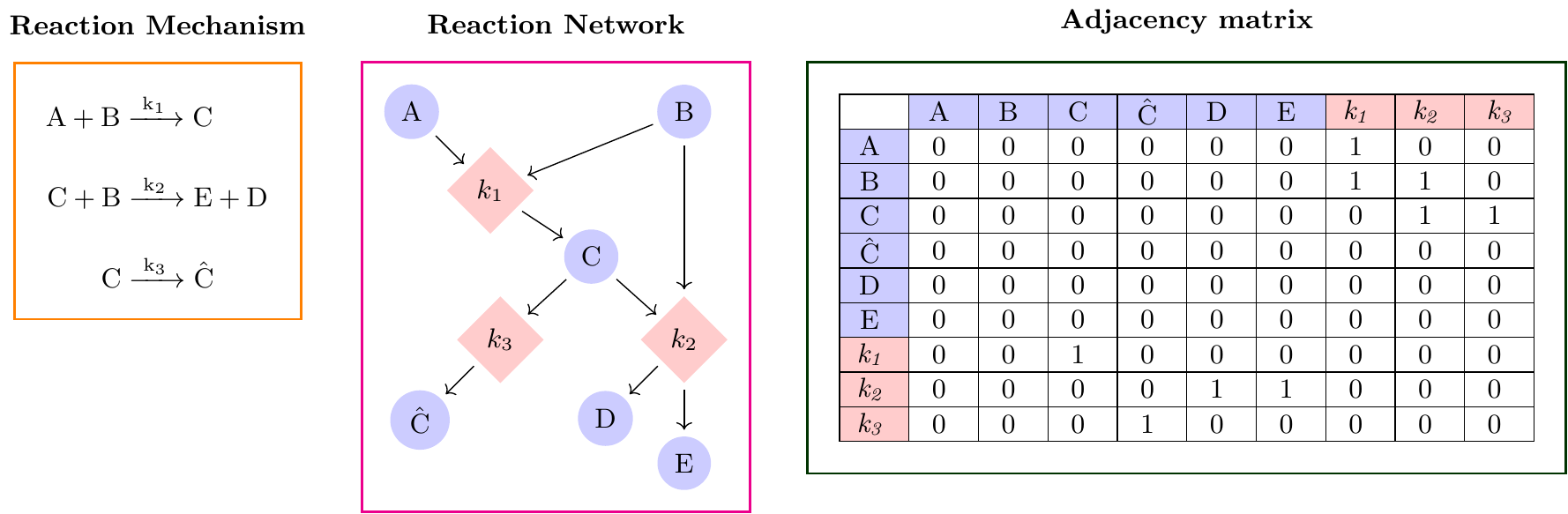}
\end{center}
\caption{Three different ways to visualize and work with a reaction network: reactions list (\textit{left panel}), colored directed graph (or colored di-graph), where circles correspond to chemical species, while squares indicate rate constants (\textit{middle panel}), and non symmetric adjacency (or connectivity) matrix representation of the colored di-graph (\textit{right panel}).}
\label{fig:graph_net}
\end{figure*}

Due to the importance to have accurate reaction energies for our reactions network cleaning methodology, we  report a benchmark of our resulting ZPE-corrected reaction energy with respect to other published studies in Table~\ref{tab:reactions_bench}.
All differences are smaller than 4 kJ/mol,  which corroborates the accuracy of present results.
Please note that the differences are mostly due to a different treatment of the spin contamination, which we improved in the present study.

We decided not to compute the reaction energies using the standard enthalpy of formation ($\Delta$H$_f^\circ$(T), usually given at 298.15 K or 0 K), but instead we used the absolute electronic energy.
This choice minimizes the propagation error in the QM computations, which would be larger if carried out using the formation enthalpy.
One way to overcome this limitation would be to compute the $\Delta$H$_f^\circ$(T) by the Isodesmic and Homodesmic cycles \citep{cramer2013essentials,wheeler2009hierarchy}, which compensates the errors. 
The problem with this technique is the difficulty of automatize it. 
In the past, some attempts were carried out but they cannot be generalized to all chemical species \citep[\textit{e.g.}][]{cavallotti2022automation}.

It is worth noticing that many astrochemical species, radicals and exotic species with respect to the terrestrial chemistry, lack laboratory measured $\Delta$H$_f^\circ$(T).
Therefore, limiting the control on the exo/endo thermicity of the reaction to only those having laboratory experimental data would greatly reduce our capacity to "clean" the astrochemical networks.

Overall, we characterized 542 species.
The whole list is publicly available at the ACO (Astro-Chemical Origin) project website\footnote{\url{https://aco-itn.oapd.inaf.it/home}}. 
In addition, a data frame (in \texttt{.csv} format) with all the properties of the studied species is available in the Supplementary Material of this article (see Table \ref{tab:support_material}.

\subsection{Graph theory for reaction networks }
\label{subsec:graph_theory}

In the astrochemical community, the reaction network is usually encoded and manipulated with a matrix approach. 
In this work, we adopt an approach based on the \textit{graph theory}, so that the network manipulation can be easily obtained via the \textsc{NetworkX} \citep{SciPyProceedings_11} Python package (\textit{\textit{e.g.}}, node connectivity, centrality, sinks, and other characteristics). 
Instead of using a multi-di-graph approach as done by \cite{grassi2013chemical}, the mathematical object we used to describe the reaction network is the colored directed graph (also known as colored di-graph).
Here, each node has a color attribute and the connections are directed \citep{barabasi2013network}. 
A colored node can have different attributes depending on the color: in our case, it can be a species-node or a rate constant-node. 
In addition, in a directed graph, the edges have a direction, \textit{i.e.} the adjacency (or connectivity) matrix is not symmetrical, so that the connection between nodes is orientated \citep[\textit{e.g.}][]{barabasi2013network,newman2018networks}. 
In Figure~\ref{fig:graph_net}, we report the codification of a generic reaction network given in the above-mentioned three ways: the simple reaction list, the reaction network visualization and, finally,  the corresponding adjacency matrix of the colored directed graph. 
The Python scripts used to encode and characterize the network properties are publicly available in a GitHub repository\footnote{\url{https://github.com/TinacciL/GreToBaPe_Cleaning}}.

\subsection{Network cleaning procedure}\label{subsec:Meth-cleaning}

\subsubsection{Identification of endothermic reactions }\label{subsubsec:cleaning-method_endo}
As explained in the Introduction, the first goal of the present  work is to produce a network "cleaned" from endothermic reactions that have not been reported as such (see below), since they cannot occur in the cold ISM.
To this end, we computed the reaction enthalpy $\Delta \mathrm{H}$ of the neutral-neutral, ion-neutral, and ion-ion reactions present in the reaction network.
Note that we did not consider photo- and electron- induced reactions (involving electrons, cosmic-ray particles and UV radiation) because they entail strongly energetic processes, whose products are usually not in their fundamental electronic state.

In the ISM, molecules primarily reside in their ground electronic state and lowest vibrational level, with only the rotational levels populated, depending on the gas conditions. 
Since the rotational energy has a negligible contribution to the total reaction energy, we used the electronic energy and the vibrational energy at 0 K (also called Zero-point Energy correction, \textit{i.e.}, ZPE) to evaluate the reaction enthalpy $\Delta \mathrm{H}(0)$ of each reaction present in the reaction network.
Then, an endothermic reaction has $\Delta \mathrm{H}(0)$ larger than zero.

In addition, in the astrochemical networks the rate constants of the neutral-neutral and ion-neutral (but not ion-polar) reactions are reported following the modified Arrhenius equation \citep{kooij1893zersetzung,laidler1996glossary}:
\begin{equation}\label{eq:Arrhenius}
k(\mathrm{T}) = \alpha \left(\frac{\mathrm{T}}{{\rm 300 ~K}}\right)^\beta \exp \left( -\frac{\gamma}{\mathrm{T}}\right)
\end{equation}
where T is the gas temperature in K, and $\alpha$, $\beta$, and $\gamma$ are often derived as fitted parameters of either experimental or theoretical values of $k(\mathrm{T})$.
It should be emphasized that $\gamma$ does not always have an actual physical meaning.
In other words, $\gamma$ is often not a real activation barrier.
That said, it can be safely assumed that, if the reaction has a barrier, it is lower or equal to $\gamma$.
Therefore, in order to consider that a reaction has been incorrectly inserted in the reaction network, we apply the additional criterion that $\Delta \mathrm{H}(0) > \gamma ~R$, where $\gamma$ is multiplied by the constant $R$ to have it in kJ/mol.

Finally, our computations have an intrinsic uncertainty, \textit{i.e.} the accuracy of the electronic structure QM calculations, which we conservatively evaluate to be 10 kJ/mol.
Therefore, when retaining or excluding a reaction from the network we consider a 10 kJ/mol threshold regardless the actual meaning of $\gamma$ either a real activation barrier or the results of an analytical fit.

In summary, a reaction is considered \textit{correctly} included in the network if it satisfies the following criteria:
\begin{itemize}
    \item [1)] if the reaction is not encoded with the modified Arrhenius equation (\textit{i.e.} the ion-pol formula for barrierless reactions), $\Delta \mathrm{H}(0) \leq 10$ kJ/mol;
    \item [2)] if the reaction is encoded with the modified Arrhenius equation, $\Delta \mathrm{H}(0) \leq \gamma ~R + 10$ kJ/mol.
\end{itemize}
Reactions that do not satisfy the two above criteria are excluded.

\subsubsection{The domino effect of removing endothermic reactions} \label{subsubsec:cleaning-domino}

In order for the network to be meaningful, each species must have at least one destruction route (\textit{i.e.} loss)  \textit{and} one formation route (\textit{i.e.} production).

Removing reactions from the network can affect other species if they are not formed or destroyed through other reactions than those removed.
In other words, if a species becomes a source (species with no formation routes) or a sink (species with no destruction routes) it is also removed from the network.
As a consequence, the reactions involving the removed species are also removed.

Therefore, the removal of one or more initial reactions triggers a domino effect, causing the removal of other reactions and species.
Our graph theory approach, described in \S ~\ref{subsec:graph_theory}, allows us to easily follow the domino effect and identify the reactions and species to be removed.

\subsection{Network reduction procedure} \label{subsec:Meth-reduced_network}

As stated in the Introduction, the second goal of this work is to produce a reduced network from the cleaned one, which only includes H-, C-, N-, O- and S- bearing species with at most six atoms, to be used when the larger network is unnecessary.
To this end, we used the following criteria and steps:
\begin{enumerate}
    \item[1.] We only consider C-, N-, O- and S-bearing species, plus Fe and Mg atoms, because their ions can be important positive charge carriers.
    \item[2.] We only consider molecules with six or less atoms (either reactant or product). 
    \item[3.] The removal of larger molecules may cause the domino effect, explained in \S ~\ref{subsubsec:cleaning-domino}, which can lead to the removal of other reactions and species also with less than six atoms.  
    \item[4.] Methanol is the only exception to rules 2 and 3.
    In the gas phase, it is formed by reactions involving molecules with more than 6 atoms, so it should be removed, based on rule 3.
    However, methanol is an important component of the interstellar grain mantles, a major carbon carrier, believed to be mainly formed on the grain surfaces and released into the gas phase by thermal and non-thermal effects.
    In practice, if the model contains surface reactions that form methanol, it could be injected into the gas-phase by e.g. the sublimation or sputtering of the grain mantles.
    Once in the gas-phase, methanol can react with species having less than 6 atoms (\textit{e.g.} HCO$^+$, H$_3^+$, OH etc..).
    For this reason, we keep methanol in the reduced network.
    If no methanol is produced by surface chemistry, then its abundance would be zero because no gas-phase reaction would form it so that it is not a source nor a sink.
    \item[5.] We remove all anions. 
    We had two reasons for that.
    The first one is that, usually, they are not crucial species in studies that necessitate a reduced network.
    The second one is that there is still a debate on their rate of formation \citep[\textit{e.g.}][]{Lara-Moreno2019PhysRevA,LaraMoreno2019}.
\end{enumerate}

\section{Results}
\label{sec:results}

\subsection{Original astrochemical network}\label{subsec:results-original-newtork}

The reaction network from which we start is based on the KIDA 2014 network \citep{wakelam2014_Kida_uva}, updated following studies from our and other groups, as follows:
\cite{hoyermann1981mechanism, defrees1985theoretical, meot1986ion, dobe1991kinetics, anicich2003index, Hamberg_2010_dimethyl_ether, Cheikh2012, lawson2012dissociative, fournier2014reactivity, Loison_2014_HCN, Balucani_2015_gasnet, Neufeld_2015_Sulfur, Vazart_2016_Formamide, Loison_2016_H2C3O, Urso_2016_CO, Codella_2017_SOLIS_II, Skouteris_2017_formamide, Fontani_2017_SOLISI, Balucani_2018_reaction, gao2018kinetics, Rosi_2018_SiS, Skouteris_2018_Ethanol, sleiman2018gas, Ascenzi_2019_dimethyl, Ayouz_2019_formamide, LaraMoreno2019, Skouteris_2019_dimethyl, Urso_2019_C2O, Balucani_2020, Codella_2020_SOLISV, Vazart_Cecarrelli_2020_acetaldehyde, blazquez2020gas, mancini2021computational}, and data, not yet published, courteously provided by Luca Mancini on some P-bearing reactions.
We also add the recombination reactions of F$^+$ and P$^+$ with free electrons, which are not present in the KIDA astrochemical reaction network, although the second one it is in the UMIST database. 
These data are taken from the database of \cite{badnell2006radiative} and given in the usual Arrhenius-Kooij format (Eq.~\ref{eq:Arrhenius}).

In summary, the pre-cleaned original network comprises 499 species and 7240 reactions.
The list is reported in the Supporting Material (\S~\ref{sec:support_material}: file \textsc{GRETOBAPE-pre.dat}).

Note that only 5768 of the 7240 reactions have thermodynamic data.
The remaining 1472 reactions do not either because they contain species for which the electronic structures are not available (236 reactions\footnote{Reactions involving the species Fe, Fe$^+$, C$_2$H$_7^+$, C$_9$H$_3$N$^+$ or bimolecular reactions having as a product an electron or photon.}) or they are recombination (640 reactions), or cosmic-ray particles and UV radiation (596 reactions).

\begin{figure}[!ht]
\begin{center}
\includegraphics[width=0.99\columnwidth]{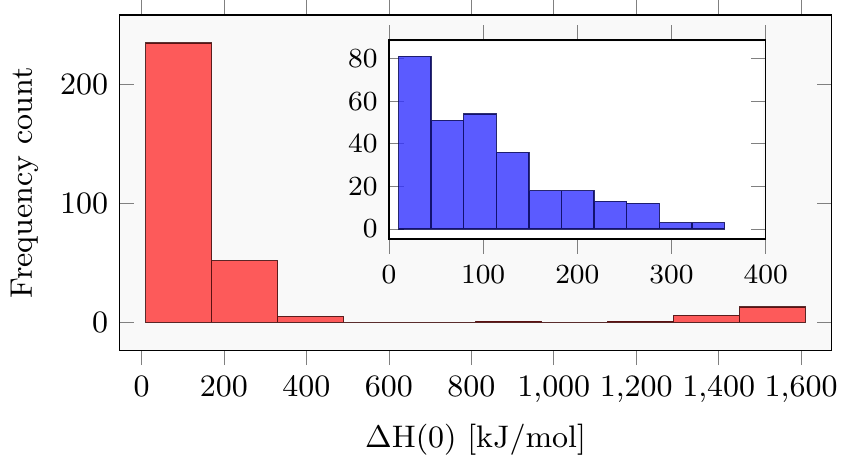}
\end{center}
\caption{Histogram of the enthalpy $\Delta \mathrm{H}(0)$ (red) of the  endothermic reactions removed from the original network after the cleaning process.
The inset (blue) shows the zoom in the 10--400 kJ/mol interval. 
The bin width is computed following the Freedman Diaconis estimator \citep{freedman1981histogram}.}
\label{fig:endo_distribution}
\end{figure}
\begin{deluxetable}{lll}[!ht]
\tablecaption{List of the species removed from the original network after the cleaning process.}
\label{tab:del_mol_info}
\tablehead{\multicolumn{3}{c}{Species}}
\startdata
SiC$_4$H        & SiC$_6$H           & SiC$_8$H           \\  
SiC$_2$CH$_3$   & SiC$_3$H$_5$       & H$_2$CSiCH         \\  
H$_2$C$_6$N$^+$ & H$_2$C$_8$N$^+$    & H$_2$C$_{10}$N$^+$ \\
PNH$_3^+$       & C$_4$H$_5$         &                    \\
\enddata
\end{deluxetable}
%

\begin{deluxetable*}{lcccccc}[!ht]
\tablecaption{Summary of the number of reactions in the original and final \textsc{GRETOBAPE} networks.}
\label{tab:net_info}
\tablehead{ & Neutral-Neutral  & Ion-Neutral & Cation-Anion & Recombination & Unimolecular & Total }
\startdata
\textbf{Pre-cleaning}   & 1007 & 3596  & 1401 & 640  & 596 & 7240 \\
\textbf{Endothermic}    & 58   & 140   & 108  & -    & -   & 306  \\
\textbf{Domino effect}  & 2    & -     & -    & 9   & 12  & 23   \\
\textbf{Post-cleaning}  & 947  & 3456  & 1293 & 631  & 584 & 6911
\enddata
\tablecomments{The reactions are listed according to their class. 
Each raw presents the number of reactions after each step of the cleaning procedure (\S ~\ref{subsec:Meth-cleaning}).  
\textit{Raw "Pre-cleaning"}: original network (\S ~\ref{subsec:results-original-newtork});
\textit{Raw "Endothermic"}: reactions removed because they are endothermic with $\Delta \mathrm{H}(0) > 10$ kJ/mol;
\textit{Raw "Domino effect"}: reactions removed because of the domino effect (\S ~\ref{subsubsec:cleaning-domino}); 
\textit{Raw "Post-cleaning"}: final cleaned network.}
\end{deluxetable*}

\subsection{New cleaned network: \textsc{GRETOBAPE}} \label{subsec:results-new-newtork}

In this section, we introduce the new reaction network obtained applying the criteria of \S ~\ref{subsec:Meth-cleaning} to the network just described in the previous section.
It is important to emphasize that we only tested the endothermicity of the original network, which is based on the products listed in KIDA and the works cited above.
We did not try to correct reactions a posteriori comparing the products in other astrochemical networks, notably UMIST (but see \S ~\ref{subsec: results-UMIST-comp}), where sometimes exothermic products of the reactions exist, for the reasons that will become clearer at the end of this section.

\subsubsection{Overview of the removed species and reactions} \label{subsec:results-overview}

\begin{deluxetable*}{c|ccccccccccc|c}[!ht]
\centering
\tablecaption{Summary of the 488 species contained in the cleaned network \textsc{GRETOBAPE}, together with the elements composing them.}
\label{tab:species_info}
\tablehead{\multirow{2}{*}{Element} & \multicolumn{11}{c|}{Number of species containing n times the element}  & \multirow{2}{*}{Total} \\ \cline{2-12}
  & \textit{\textbf{1}} & \textit{\textbf{2}} & \textit{\textbf{3}} & \textit{\textbf{4}} & \textit{\textbf{5}} & \textit{\textbf{6}} & \textit{\textbf{7}} & \textit{\textbf{8}} & \textit{\textbf{9}} & \textit{\textbf{10}} & \textit{\textbf{11}} &  }
\startdata
H  & 123  & 81 & 53 & 38 & 21 & 12 & 7  & 1  & 0  & 0  & 0 & 336 \\
He & 3    & 0  & 0  & 0  & 0  & 0  & 0  & 0  & 0  & 0  & 0 & 336 \\
C  & 100  & 66 & 57 & 34 & 21 & 19 & 20 & 17 & 19 & 12 & 2 & 367 \\
N  & 115  & 8  & 0  & 0  & 0  & 0  & 0  & 0  & 0  & 0  & 0 & 123 \\
O  & 93   & 24 & 0  & 0  & 0  & 0  & 0  & 0  & 0  & 0  & 0 & 117 \\
F  & 7    & 0  & 0  & 0  & 0  & 0  & 0  & 0  & 0  & 0  & 0 & 7   \\
Na & 6    & 0  & 0  & 0  & 0  & 0  & 0  & 0  & 0  & 0  & 0 & 6   \\
Mg & 3    & 0  & 0  & 0  & 0  & 0  & 0  & 0  & 0  & 0  & 0 & 3   \\
Si & 53   & 0  & 0  & 0  & 0  & 0  & 0  & 0  & 0  & 0  & 0 & 53  \\
S  & 38   & 7  & 0  & 0  & 0  & 0  & 0  & 0  & 0  & 0  & 0 & 45  \\
P  & 38   & 0  & 0  & 0  & 0  & 0  & 0  & 0  & 0  & 0  & 0 & 38  \\
Cl & 10   & 0  & 0  & 0  & 0  & 0  & 0  & 0  & 0  & 0  & 0 & 10  \\
Fe & 2    & 0  & 0  & 0  & 0  & 0  & 0  & 0  & 0  & 0  & 0 & 2   \\
\enddata
\end{deluxetable*}

Among the 5768 reactions that have thermodynamic data (see \S~ \ref{subsec:results-original-newtork}) and for which we could carry out the endothermicity verification, 306 do not satisfy the two criteria described in \S ~\ref{subsec:Meth-cleaning}, namely $\sim 5$\% of the studied reactions\footnote{
Please note that the reaction H + O$^+$ $\rightarrow$ H$^+$ + O is endothermic by ~17kJ/mol according our computations.
However, the reaction has been studied by other authors \citep[e.g.][]{stancil1999charge} and found to be exothermic. 
The reason for our erroneous result is that the ionization energy of O is a critical case that pushes the computational calculations to their limits and a difficult case for QM calculations due to the correlation errors even if we are using one of the best available methods.}.
Figure \ref{fig:endo_distribution} shows the distribution of their endothermic energy.
The majority of these reactions have endothermicities lower than about 150 kJ/mol while about 140 reactions even larger.

The list of the removed reactions based on the endothermicity criteria is reported in the Supporting Material (\S~\ref{sec:support_material}: file \textsc{GRETOBAPE-endo.dat}).
In addition, we supply the list of 47 reactions which are endothermic but with a reaction enthalpy lower than 10 kJ/mol and for which more accurate calculations should be carried out before removing them for the network (\S~\ref{sec:support_material}: 
file\textsc{GRETOBAPE-endo-0-10kJmol.dat}).
Finally, because of the domino effect, described in \S ~\ref{subsubsec:cleaning-domino}, 11 species and 23 additional reactions are removed, leading to a total of 329 removed reactions (\S~\ref{sec:support_material}: files \textsc{GRETOBAPE-endo.dat} plus \textsc{GRETOBAPE-domino.dat}).

The full list of deleted species is reported in Tab. \ref{tab:del_mol_info}.
Table \ref{tab:net_info} provides an overview of the removed reactions, grouped according to each reaction class.
The ion-neutral reactions are the most affected by the cleaning process in number (140), while the cation-anion reactions are the most affected in percentage ($7.7$\%).
Table \ref{tab:species_info} lists the number of species in the final cleaned network classified with their elements.

In summary, the new cleaned network \textsc{GRETOBAPE} contains 6911 reactions and 488 species.

\subsubsection{Description of the removed reactions} \label{subsubsec:cleaning-sources-errors}

The 329 removed reactions have the following properties:
\begin{itemize}
    \item 65 reactions are not encoded with the modified Arrhenius equation: they either do not satisfy the criterion 1 listed in \S ~\ref{subsubsec:cleaning-method_endo}, namely $\Delta \mathrm{H}(0) \leq 10$ kJ/mol, or they are removed because of the domino effect.
    \item 219 reactions are encoded via the Arrhenius-Kooji equation and have $\gamma=0$: they either do not satisfy the criterion 2 listed in \S ~\ref{subsubsec:cleaning-method_endo}, namely $\Delta \mathrm{H}(0) \leq 10$ kJ/mol, or they are removed because of the domino effect.
    \item 45 reactions are encoded via the Arrhenius-Kooji equation and have $\gamma \neq 0$: they either do not satisfy the criterion 2 listed in \S ~\ref{subsubsec:cleaning-method_endo}, namely $\Delta \mathrm{H}(0) \leq \gamma ~R + 10$ kJ/mol, or they are removed because of the domino effect.
\end{itemize}

Overall, the cleaning process strongly affects the Silicon chemistry, as it will be shown by the modeling simulations of \S ~\ref{subsec:astro-originalvsclean}. 
Indeed, out of a total number of 329 removed reactions, 48 involve Si-bearing species (\textsc{GRETOBAPE-domino-endo-Si}).
In practice,  about 15\% of the total removed reactions regard Si-bearing species, and about 9\% of the original reactions involving Si-bearing species (526) are removed.
Likewise, 6 out of 11 removed species are Si-bearing species (see Tab.~\ref{tab:del_mol_info}) and represents the $\sim 10$\% of the total original Si-bearing species (61).

\begin{figure*}[!ht]
    \centering
    \includegraphics[width=\textwidth]{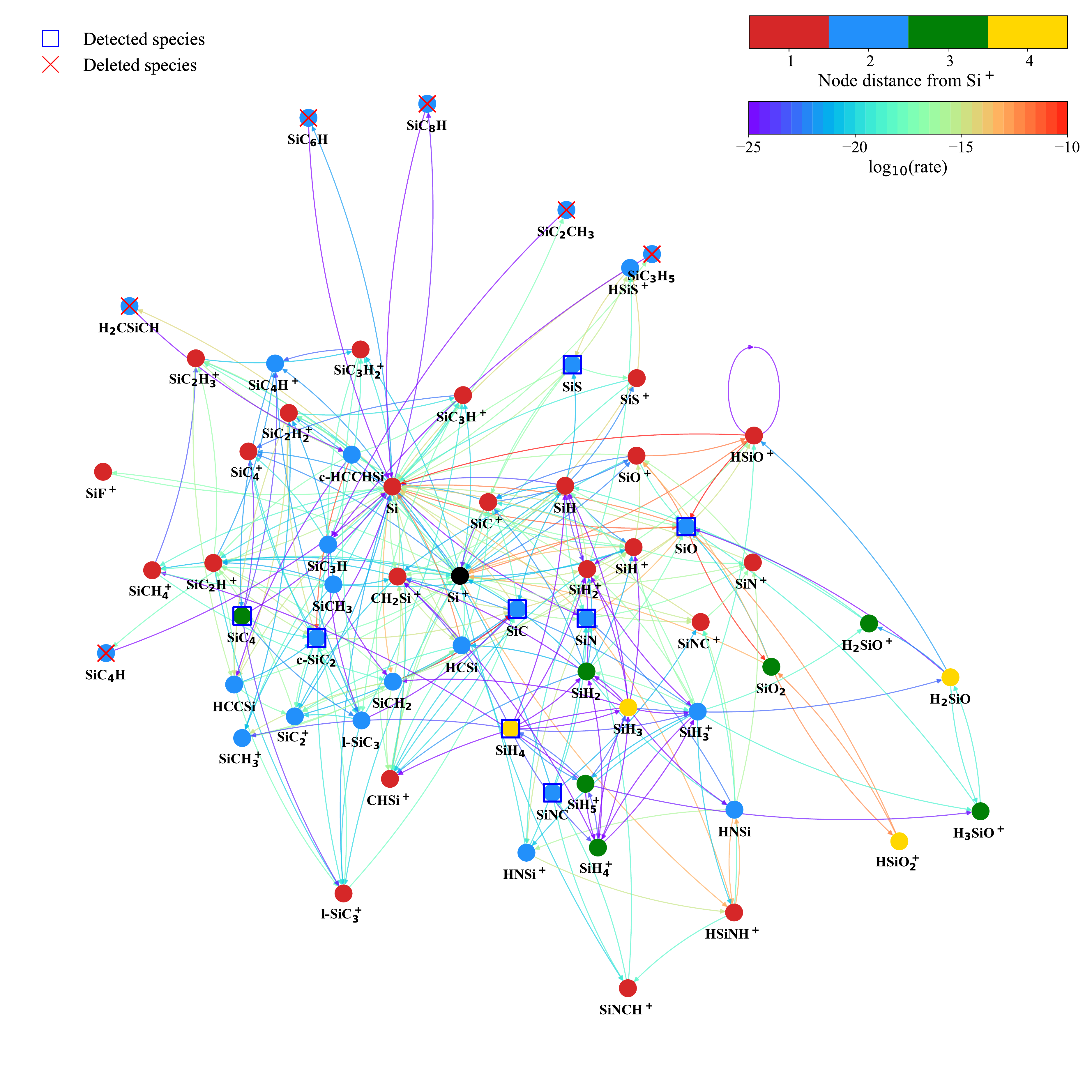}
    \caption{Original Si-bearing reaction network, visualized as directed colored graph. 
    The reactions are represented as solid lines and the species as nodes (differently from the representation described in \S ~\ref{subsec:graph_theory}, where reactions and species are both represented by nodes). 
    The lines are colored based on the \texttt{log}$_{10}$ of the reaction rate constants calculated at 90 K and multiplied by the reactants densities at $5\times10^3$yr in the "warm molecular outflow shock" model described in \S ~\ref{subsubsec:astro-shock}, using the original (pre-cleaning) network.
    All rates larger than $10^{-10}$ are red and those less than $10^{-25}$ are violet.
    The nodes are colored based on the graph distance of that species from Si$^+$ (highlighted in black).
    The species removed from the original network are marked by red crosses, while those detected in the ISM by blue squares. 
    Please note that the digital format of the figure is vectorial and can be zoomed to better look at the details.}
    \label{fig:Si_network}
\end{figure*}
\begin{figure*}[!ht]
    \centering
    \includegraphics[width=\textwidth]{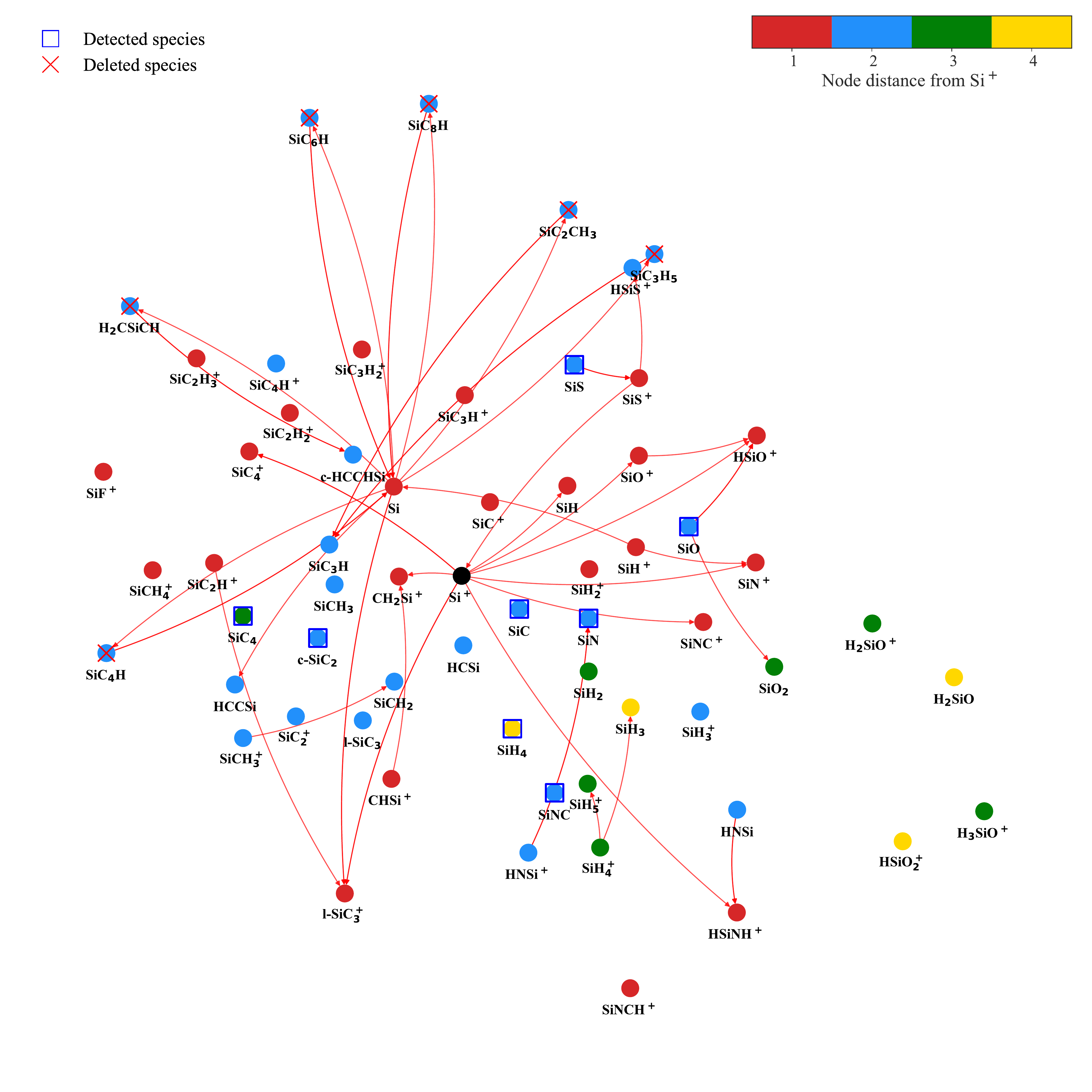}
    \caption{ As Fig.~\ref{fig:Si_network}, showing only the reactions removed from the cleaning process.}
    \label{fig:Si_deleted_network}
\end{figure*}

Figure \ref{fig:Si_network} shows a graph representation of the reactions involving Si-bearing species and  present in the original network, together with the removed and detected species. 
Figure \ref{fig:Si_deleted_network} presents the same reaction network showing the deleted reactions only.
First, the 6 removed Si-bearing species (all containing more than 6 atoms) are only formed and deleted by one reaction, so that their removal did not introduce a domino effect on other species. 
As can be seen from Figs. \ref{fig:Si_network} and \ref{fig:Si_deleted_network}, these species are in fact at the border of the reaction network, meaning that they have only a minor role in the overall Si reaction network.
Second, all the removed reactions involving Si-bearing species have endothermic formation routes.

Digging into the KIDA database to understand their literature origin, we find that almost all these reactions are introduced not because of specific experiments or calculations, but based on educated guesses. 
Specifically, their rate constants are estimated via the capture theory approach described in \cite{Herbst2006} and \cite{woon2009quantum} for ion-neutral reactions. 
For example, the formation route of SiC$_4$H, SiC$_6$H and SiC$_8$H is assumed to be Si + C$_x$H$_2$ $\rightarrow$ SiC$_x$H + H, and their constant rates are estimated.

Furthermore, looking in detail at the deleted reactions involving Si-bearing species (Fig. \ref{fig:Si_deleted_network}) most of them are reactions that increase the complexity of the molecules, namely they increase their number of atoms. 
This is a general characteristic of the cleaning process: the most affected class of reactions are those that increase the number of atoms of at least one reactant rather than those that reduce it. 
This result can be explained by the fact that the excess of energy of "destructive" reactions is generally much easier to guess than that of formation reactions.

Another source of endothermic reactions comes from the cation-anion reactions which involve carbon chain anions, whose data were predominantly taken from \cite{harada2008modeling}.

\paragraph{Final remarks} It is of paramount importance to emphasize that the network \textsc{GRETOBAPE} is the result of the removal of endothermic reaction channels.
Consequently, some loss/production channels have disappeared from the network, some of which are essential for the production of detected interstellar species.
The list of removed reactions needs, therefore, a dedicated systematic (and time-consuming) revision as will be shown in \S ~\ref{subsec: results-UMIST-comp}.

\subsubsection{Comparison with the UMIST database}\label{subsec: results-UMIST-comp}

As already emphasized at the beginning of \S ~\ref{subsec:results-new-newtork}, the original network on which we tested the endothermicity of the reactions is based on the KIDA 2014 network and, therefore, the 329 reaction channels removed refer to the KIDA database. 
In this subsection, we discuss some deleted reactions in comparison with the data reported in UMIST, in which sometimes a more detailed bibliography is reported.
It is also worth mentioning that our scope here is not to substitute the precious work done by databases maintainers, but rather to highlight some identified problems.

\paragraph{Reaction Si + C$_2$H$_2$ $\rightarrow$ SiC$_2$H + H}
UMIST reports two products, SiC$_2$ + H$_2$ and SiC$_2$H + H, with the second one three times faster than the first.
They assumed this branching ratio based on the study at 15 K by \cite{canosa2001rate} of the global rate, and based on \cite{kaiser2009chemical} for the products.
However, the analysis by \cite{kaiser2009chemical} shows that the SiC$_2$H + H is indeed endothermic.

\paragraph{Reaction S$^+$ + CH$_4$ $\rightarrow$ H$_3$CS$^+$ + H}
This reaction is crucial for the formation of the organo-sulphur products, and we deleted it since endothermic, despite it is considered as barrierless in KIDA and UMIST (with the same parameters). 
UMIST reports the reference of the experimental data by \cite{smith1981reactions}. 
These authors assumed that H$_3$CS$^+$ is formed, which is impossible because S$^+$ is a quadruplet while H$_3$CS$^+$ is a triplet and it would be impossible energetically for the spin conservation rules. 
Indeed, \cite{yu2020electronic} found that the corrected product seen by \cite{smith1981reactions} is H$_2$CSH$^+$ (singlet) and not of H$_3$CS$^+$, as corroborated by our calculations that found the channel producing H$_2$CSH$^+$ to be exothermic.
Please note that to re-introduce the reaction with the correct reaction channel (H$_2$CSH$^+$) would require to revise all the reactions involving H$_3$CS$^+$ and H$_2$CSH$^+$, which is beyond the scope of the present work.


\paragraph{Reaction CN$^+$ + CO$_2$ $\rightarrow$ NO + C$_2$O$^+$} 
Experiments sometimes provided wrong products, as in the case of the channel CN$^+$ + CO$_2$ $\rightarrow$ NO + C$_2$O$^+$ which is reported in both KIDA and UMIST and taken from the experimental work by \cite{raksit1984selecte}. 
However, this channel is endothermic by more than 138 kJ/mol, as found by our computations (which are also validated by thermochemical data\footnote{\url{https://atct.anl.gov/}}) considering either of the two linear isomers, CCO$^+$ and COC$^+$. 
In this case, therefore, the problem is not due to the possible isomer structure but to the experiment itself.
Indeed, another experiment by \cite{mcewan1983reactions} did not find the NO + C$_2$O$^+$ channel but two other channels, also found by \cite{raksit1984selecte}, \textit{i.e.} CN + CO$_2$+ and CO + NCO$_2$, both of them exothermic.

\subsubsection{Missing destruction reactions} \label{par:missing_destruction}

In cold molecular clouds, with few exceptions, neutral species are mainly destroyed by the most abundant cations, namely H$_3^+$, HCO$^+$, H$_3$O$^+$, He$^+$ and H$^+$. 
Therefore, every neutral species should have at least one destruction channel for each of the above cations. 
Table \ref{tab:missing_distruction} lists all the neutral species having no destruction channels with the aforementioned cations in the cleaned reaction network. 
Please note that some reaction channels were removed during the cleaning process.
We mark them with a ``$\bullet$" symbol. 
The Tab. \ref{tab:missing_distruction} list is meant to identify the reactions that need experiments and/or QM computations.
In addition, our database of absolute electronic energies, ZPE and $m_s$ \citep{ISM_cations} can be used to test the possible reaction channels of formation and/or destruction routes. 

As an illustrative example, it is worth mentioning the destruction reactions involving SiO. 
The reaction of SiO with H$_3$O$^+$ $\rightarrow$ H$_2$O +	HSiO$^+$ has an enthalpy of +152.1 kJ/mol. 
A similar endothermicity is found for the reaction SiO + HCO$^+$ $\rightarrow$ CO +	HSiO$^+$.
However, if the product of both reactions is not HSiO$^+$, as reported in the KIDA database, but its constitutional isomer SiOH$^+$, as reported in UMIST, the reactions become respectively exothermic of -112.2 and -208.7 kJ/mol. 
However, since in KIDA HSiO$^+$ is produced from H$_2$SiO$^+$, which is in turn produced from H$_3$SiO$^+$, and HSiO$^+$  is involved in a dozen reactions (see Fig. \ref{fig:Si_network}), a simple editing may lead to wrong results.
A more detailed study on the reliability of the SiOH$^+$ isomer as a product is postponed to a future work by our group focusing on the improvement of the Silicon reaction network (clearly necessary, as highlighted in \S ~\ref{subsubsec:cleaning-sources-errors}).


\subsubsection{Missing formation reactions}

The aim of the present work is to provide a reaction network as reliable as possible.
Identifying important missing reactions is, therefore, an aspect of this goal.
In addition to the missing destruction reactions described in the previous subsection, here we list obvious missing formation reactions, namely those forming detected species. To this scope, we used the recent \citet{McGuireCensus2021} census of the detected species in the ISM and extragalactic medium. 
Table~\ref{tab:missing_species} reports the detected species which do not have a formation route in our cleaned network \textsc{GRETOBAPE}. 
Most of these molecules are also missing from commonly used databases (\textit{e.g.}, KIDA). 
Hopefully, Tab. \ref{tab:missing_species} can be used to motivate future works and strengthen the interdisciplinary collaboration between chemists and astronomers.

\subsection{Reduced network: \textsc{GRETOBAPE-red}} \label{subsec:results-red-network}
Following the criteria described in Section~\ref{subsec:Meth-reduced_network}, the reduced network \textsc{GRETOBAPE-red} contains 204 chemical species from the 488 existent in the (cleaned) network \textsc{GRETOBAPE}. 
In addition, the removal of reactions associated with pruned species reduces the size of the network from 7240 reactions to 2810. 
Compared to \textsc{GRETOBAPE}, \textsc{GRETOBAPE-red} has less than half the amount of reactions and species, implying a significant decrease in computation time when used in any chemical code. 
This performance boost is seen in the tests we performed on the reduced network using the \textsc{krome} package (see next section, \S~\ref{sec:implications}), where the computing time of a typical gas-phase run decreases from 14 to 3 s using the complete and the reduced networks, respectively.

Due to the criteria imposed on the creation of this network, the main discrepancy we expect between the complete and the reduced networks concerns the chemistry of medium-sized (between 3-6 atoms) carbon chains. 
This is due to removing potentially important destruction routes, \textit{i.e.} reactions between these medium-sized carbon chains to form larger ones when pruning out C-bearing species with more than 6 atoms. 
Nonetheless, a complete assessment of the reliability of this reduced network is discussed in the following section, following detailed modeling of a cold and warm case, respectively.

\begingroup
\setlength{\tabcolsep}{2.5pt} 
\startlongtable
\begin{deluxetable*}{lccccc|lccccc|lccccc}
\tablecaption{List of neutral species with missing reactions with the most abundant cations in the cleaned newtork \textsc{GRETOBAPE}.}
\tablehead{ Species &  H$^+$ &  He$^+$ &  H$_3^+$ &  H$_3$O$^+$ &  HCO$^+$  & Species  &  H$^+$ &  He$^+$ &  H$_3^+$ &  H$_3$O$^+$ &  HCO$^+$ & Species  &  H$^+$ &  He$^+$ &  H$_3^+$ &  H$_3$O$^+$ &  HCO$^+$}
\label{tab:missing_distruction}
\startdata
NH               &            &            &            &  $\star$    &            &   HPO              &            &            &            &  $\star$   &           &   CH$_2$CHCN       &            &            &            & $\star$     &            \\
OH               &            &            &            &  $\star$    &            &   OCS              &            &            &            & $\star$    &           &   C$_6$H           &            &            &            & $\star$     &            \\
HF               &  $\star$   &            &            &  $\star$    &  $\star$   &   SiO$_2$          &  $\star$   &            &            & $\star$    &           &   HC$_5$N          &            &            &            &  $\star$    &            \\
NaH              &            &            &            &  $\star$    &            &   SO$_2$           &  $\star$   &            &            & $\star$    & $\star$   &   C$_7$            &            &            &            &  $\star$    &            \\
C$_2$            &            &            &            &  $\star$    &            &   CH$_3$           &            &            &            & $\star$    &           &   C$_6$N           &  $\star$   &  $\star$   &            &  $\star$    &            \\
MgH              &            &            &            &  $\star$    &            &   H$_2$CN          &  $\star$   &  $\star$   &  $\star$   & $\star$    & $\star$   &   C$_2$H$_6$       &            &            &            &  $\star$    &            \\
CN               &            &            &            &             &  $\star$   &   SiH$_3$          &            &            &            & $\star$    & $\star$   &   C$_3$H$_5$       &  $\star$   &  $\star$   &  $\star$   &  $\star$    &  $\star$   \\
CO               &  $\star$   &            &            &  $\star$    &  $\star$   &   HOOH             &  $\star$   &  $\star$   &  $\star$   & $\star$    & $\star$   &   CH$_3$OCH$_2$    &  $\star$   &            &  $\star$   &             &            \\
N$_2$            &  $\star$   &            &            &  $\star$    &            &   SiCH$_2$         &            &            &            & $\star$    &           &   CH$_2$CHC$_2$H   &  $\star$   &  $\star$   &  $\star$   &  $\star$    &  $\star$   \\
NO               &            &            &            &  $\star$    &  $\star$   &   H$_2$CS          &            &            &            & $\star$    &           &   CH$_3$C$_3$N     &            &            &            &  $\star$    &            \\
PH               &            &            &            &  $\star$    &            &   C$_3$N           &  $\star$   &            &            &            &           &   C$_7$H           &            &            &            &  $\star$    &            \\
O$_2$            &            &            &            &  $\star$    &            &   l-SiC$_3$        &            &            &            & $\star$    &           &   HC$_6$N          &  $\star$   &            &  $\star$   &  $\star$    &  $\star$   \\
HS               &            &            &            &  $\star$    &            &   C$_3$P           &  $\star$   &            &            &            &           &   C$_8$            &            &            &            &  $\star$    &            \\
HCl              &            &            &            &  $\star$    &  $\star$   &   CH$_4$           &            &            &            & $\star$    &           &   C$_7$N           &  $\star$   &            &            &  $\star$    &            \\
SiN              &            &            &            &  $\star$    &  $\star$   &   CH$_2$NH         &            &            &            & $\star$    &           &   C$_7$O           &  $\star$   &  $\star$   &            &  $\star$    &            \\
SiO              &            &            &            & $\bullet$   & $\bullet$  &   CH$_3$O          &  $\star$   &            &            & $\star$    &           &   CH$_3$CHCH$_2$   &  $\star$   &  $\star$   &  $\star$   &  $\star$    &  $\star$   \\
NS               &            &            &            &  $\star$    &            &   SiH$_4$          &            &            &            & $\star$    &           &   CH$_3$C$_4$H     &            &            &            &  $\star$    &            \\
PO               &            &            &            &  $\star$    &            &   H$_2$CCN         &            &            &            & $\star$    & $\star$   &   C$_8$H           &            &            &            &  $\star$    &            \\
CCl              &  $\star$   &            &  $\star$   &  $\star$    &  $\star$   &   NH$_2$CN         &            &            &            & $\star$    &           &   HC$_7$N          &            &            &            &  $\star$    &            \\
ClO              &  $\star$   &            &  $\star$   &  $\star$    &  $\star$   &   CH$_2$PH         &            &            &            & $\bullet$  & $\bullet$ &   C$_9$            &            &            &            &  $\star$    &            \\
SO               &            &            &            &  $\star$    &            &   HCOOH            &            &            &            & $\star$    &           &   C$_8$N           &  $\star$   &  $\star$   &            &  $\star$    &            \\
SiS              &            &            &            &  $\star$    &            &   c-HCCHSi         &            &            &            & $\star$    &           &   C$_3$H$_7$       &  $\star$   &  $\star$   &  $\star$   &  $\star$    &  $\star$   \\
S$_2$            &            &            &            & $\bullet$   &            &   C$_5$            &            &            &            & $\star$    &           &   CH$_2$CHCHCH$_2$ &  $\star$   &  $\star$   &  $\star$   &  $\star$    &  $\star$   \\
H$_2$O           &            &            &            &  $\star$    &            &   C$_4$N           &  $\star$   &  $\star$   &            & $\star$    &           &   CH$_3$C$_5$N     &            &            &            &  $\star$    &            \\
HCO              &            &            &            &  $\star$    &            &   SiC$_3$H         &            &            &            & $\star$    &           &   C$_9$H           &            &            &            &  $\star$    &            \\
HNO              &            &            &            &  $\star$    &            &   SiC$_4$          &            &            &            & $\star$    &           &   HC$_8$N          &  $\star$   &            &  $\star$   &  $\star$    &  $\star$   \\
PH$_2$           &            &            &            &  $\star$    &            &   HCCCHO           &  $\star$   &  $\star$   &            &            &           &   C$_{10}$         &            &            &            &  $\star$    &            \\
O$_2$H           &  $\star$   &  $\star$   &  $\star$   &  $\star$    &  $\star$   &   c-C$_3$H$_2$O    &  $\star$   &  $\star$   &            &            &           &   C$_9$N           &  $\star$   &            &            &  $\star$    &            \\
CCN              &            &            &            &  $\star$    &            &   H$_2$CCCO        &  $\star$   &  $\star$   &            & $\bullet$  &           &   C$_9$O           &  $\star$   &  $\star$   &            &  $\star$    &            \\
NaOH             &            &            &            &  $\star$    &            &   C$_5$H           &            &            &            & $\star$    &           &   C$_3$H$_8$       &  $\star$   &  $\star$   &  $\star$   &  $\star$    &  $\star$   \\
CCO              &            &            &            & $\bullet$   &            &   HC$_4$N          &  $\star$   &            &  $\star$   & $\star$    & $\star$   &   CH$_3$C$_6$H     &            &            &            &  $\star$    &            \\
OCN              &  $\star$   &            &  $\star$   &  $\star$    &  $\star$   &   C$_6$            &            &            &            & $\star$    &           &   C$_{10}$H        &            &            &            &  $\star$    &            \\
HNSi             &            &            &            & $\bullet$   &            &   C$_5$N           &  $\star$   &            &            & $\star$    &           &   HC$_9$N          &            &            &            &  $\star$    &            \\
CO$_2$           &            &            &            &  $\star$    &            &   C$_5$O           &  $\star$   &  $\star$   &            & $\star$    &           &   C$_{11}$         &  $\star$   &            &  $\star$   &  $\star$    &  $\star$   \\
N$_2$O           &            &            &            &  $\star$    & $\bullet$  &   C$_2$H$_5$       &            &            &            & $\star$    &           &   C$_{10}$N        &  $\star$   &  $\star$   &            &  $\star$    &            \\
HCS              &            &            &            &  $\star$    &  $\star$   &   CH$_3$NH$_2$     &            &            &            & $\star$    &           &   c-C$_6$H$_6$     &  $\star$   &            &            &       &      \\ 
NO$_2$           &  $\star$   &  $\star$   &            &  $\star$    &  $\star$   &   CH$_2$CCH$_2$    &  $\star$   &  $\star$   &  $\star$   & $\star$    & $\star$   &   CH$_3$C$_7$N     &            &            &            &  $\star$    &            
\enddata
\tablecomments{With the ``$\bullet$'' and ``$\star$'' symbols we identify the neutral species reaction with the corresponding cation that was present in our network but was deleted in cleaning process, and the reaction that was originally absent, respectively.}
\end{deluxetable*}
\endgroup
%

\begin{deluxetable*}{llllllll}
\tablecaption{Species with reported observations in the ISM \citep{McGuireCensus2021} which are not present in the final \textsc{GRETOBAPE} network. 
}
\tablehead{\multicolumn{7}{c}{\textbf{Carbon, nitrogen and oxygen-bearing species}}}
\label{tab:missing_species}
\startdata
\multicolumn{7}{l}{\textit{4-atoms}} \\
HCCO & HONO & HNCN & CNCN & HC$_2$N & & \\
\multicolumn{7}{l}{\textit{5-atoms}} \\
HNCNH & NH$_2$OH & H$_2$NCO$^+$ & NCCNH$^+$ & & & \\
\multicolumn{7}{l}{\textit{6-atoms}} \\
HNCHCN & CH$_3$NC & C$_5$N$^-$ & & & & \\
\multicolumn{7}{l}{\textit{7-atoms}} \\
c-C$_2$H$_4$O & c-C$_3$HCCH & CH$_2$CHOH & HC$_4$NC & CH$_3$NCO & HOCH$_2$CN & HC$_5$O \\
\multicolumn{7}{l}{\textit{8-atoms}} \\
CH$_2$CHCHO & CH$_3$CHNH & NH$_2$CH$_2$CN & NH$_2$CONH$_2$ & & & \\
\multicolumn{7}{l}{\textit{9-atoms}} \\
HC$_7$O & H$_2$CCCHCCH & HCCCHCHCN & H$_2$CCHC$_3$N & CH$_3$NHCHO & CH$_3$CONH$_2$ & \\
\multicolumn{7}{l}{\textit{10-atoms}} \\
HOCH$_2$CH$_2$OH & CH$_3$CH$_2$CHO & CH$_3$CHCH$_2$O & CH$_3$OCH$_2$OH & & & \\
\multicolumn{7}{l}{\textit{11-atoms}} \\
C$_2$H$_5$OCHO & CH$_3$COOCH$_3$ & CH$_3$COCH$_2$OH & NH$_2$CH$_2$CH$_2$OH & & & \\
\multicolumn{7}{l}{\textit{12-atoms}} \\
n-C$_3$H$_7$CN & i-C$_3$H$_7$CN & 1-C$_5$H$_5$CN & 2-C$_5$H$_5$CN & & & \\
\multicolumn{7}{l}{\textit{$>$13-atoms}} \\
HC$_{11}$N & c-C$_6$H$_5$CN & C$_9$H$_8$ & 1-C$_{10}$H$_7$CN & 2-C$_{10}$H$_7$CN & & \\
\hline
\multicolumn{7}{c}{\textbf{Sulfur-bearing species}} \\
\hline
HSC & NCS & HNCS & HCCS & CHOSH & HCSCN & HCSCCH \\
H$_2$CCS & H$_2$CCCS & CH$_3$SH & C$_5$S & CH$_3$CH$_2$SH & & \\
\hline
\multicolumn{7}{c}{\textbf{Silicon-bearing species}} \\
\hline
SiCN & SiCSi & SiC$_3$ & SiH$_3$CN & CH$_3$SiH$_3$ & &  
\enddata
\tablecomments{Please note that different isomers of the listed species may be present in KIDA and UDfA.}
\end{deluxetable*}

\section{Astrochemical implications}
\label{sec:implications}
This section illustrates the impact of the new reaction networks on the predicted species abundances, obtained using the original reaction network versus the new cleaned \textsc{GRETOBAPE} and the reduced \textsc{GRETOBAPE-red} networks, respectively.
To this end, we run two time-dependent gas-only astrochemical codes:
\textsc{MyNahoon}, a modified version of the publicly available code \textsc{Nahoon} \citep{wakelam2012kinetic}\footnote{Briefly, the modifications concern a more friendly usage of the code, in input and output, and not the core of the code itself.}, 
and \textsc{krome}, a publicly available package \citep{Grassi2014}.
\textsc{MyNahoon} is used to evaluate the difference in the species abundances predicted with the original and the new \textsc{GRETOBAPE} reaction network in a typical cold molecular cloud (\S ~\ref{subsubsec:astro-cloud}) and a typical warm molecular outflow shock (\S ~\ref{subsubsec:astro-shock}), respectively.
\textsc{krome} is used to test the reliability of the reduced network \textsc{GRETOBAPE-red} (\S ~\ref{subsec:astro-reduced-network}) in a large parameter space.

\subsection{Predicted abundances with the original versus the cleaned reaction network \textsc{GRETOBAPE} }\label{subsec:astro-originalvsclean}

In order to assess the difference between the obtained clean reaction network \textsc{GRETOBAPE} and the original one, we performed two simulations: the first one simulates a typical cold molecular cloud (\S ~\ref{subsubsec:astro-cloud}) and the second one a typical warm molecular outflow shock (\S ~\ref{subsubsec:astro-shock}).

\subsubsection{Cold molecular cloud}
\label{subsubsec:astro-cloud}

\begin{figure*}
    \caption{Comparison of the network post- (\textsc{GRETOBAPE}) over pre-cleaning predicted abundance ratios (left panels) and post-cleaning abundance (right panels) as a function of time for the cold molecular cloud model (\S ~\ref{subsubsec:astro-cloud}). 
    The three rows show three groups of species where the abundance ratios during the evolution are $\geq10^3$ or $\leq10^{-3}$ (upper panels), 10--$10^3$ or 0.1--$10^{-3}$ (middle panels) and 3--10 or 0.1--0.3 (bottom panels). 
    Only species with abundances larger than 10$^{-13}$ at steady-state are considered.}
    \label{fig:comparison_cloud}
\begin{center}
    \includegraphics[width=0.99\textwidth]{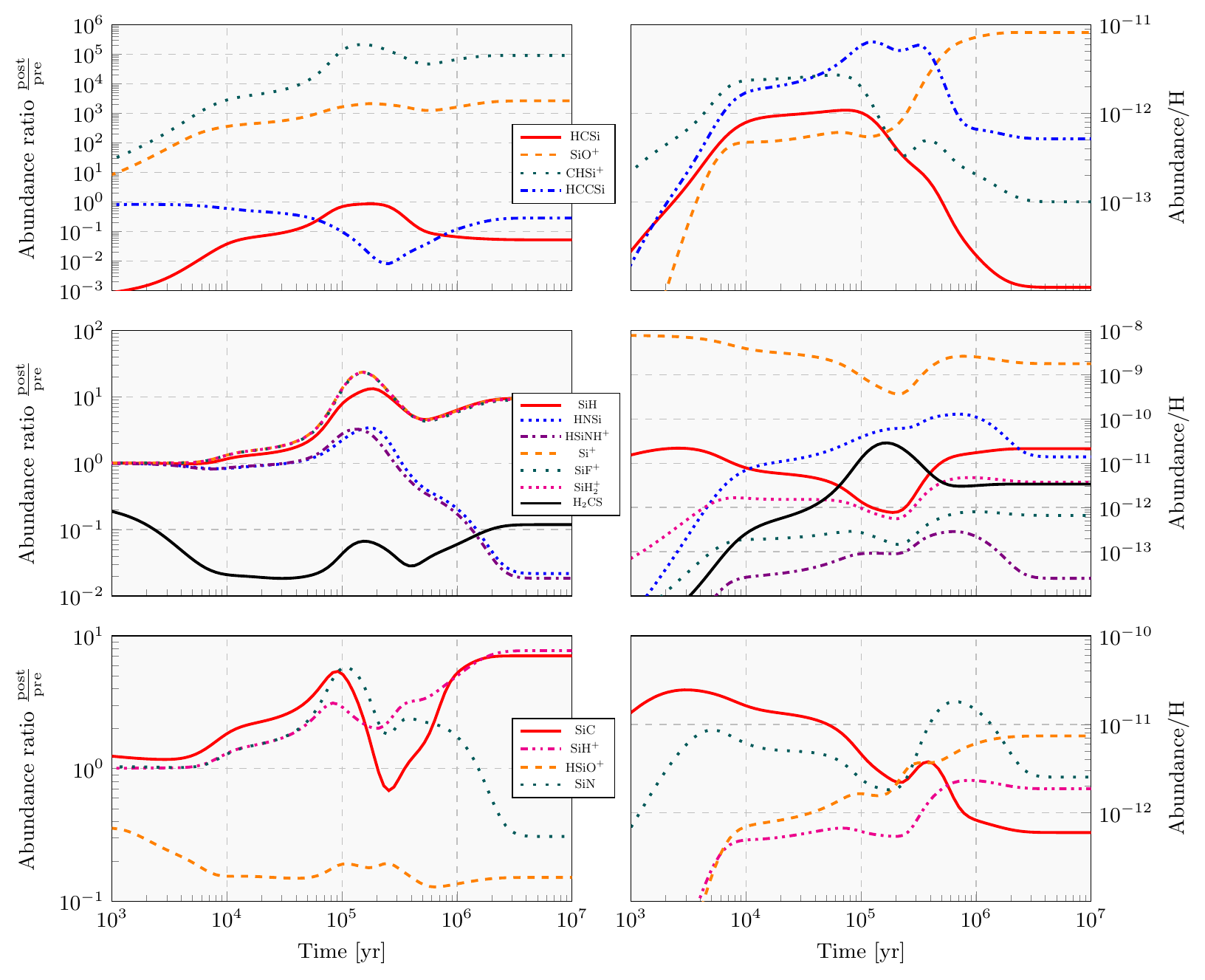}
\end{center}
\end{figure*}

We adopted the parameters of a typical molecular cloud, namely: 
temperature (gas and dust) equal to 10 K, H nuclei number density equal to $2 \times 10^4$ cm$^{-3}$, visual extinction A$_v$ equal to 20 mag, cosmic-ray (CR) ionization rate for H$_2$ $\zeta$ equal to $3 \times 10^{-17}$ s$^{-1}$.
The initial elemental abundances are as follows \citep[from][with the heavy metals depleted by a factor 100]{jenkins2009unified}:
He/H = $9.0\times 10^{-2}$, C/H = $2.0\times 10^{-4}$, O/H = $2.6\times 10^{-4}$, N/H = $6.2\times 10^{-5}$, S/H = $8\times 10^{-8}$, Si/H = $8\times 10^{-9}$, Fe/H = $3\times 10^{-9}$, Mg/H = $7\times 10^{-9}$.
Finally, we started with the conditions of a pseudo-translucent cloud, namely with hydrogen in the molecular form and the other elements in atomic (ionic or neutral) form, and we let evolve the chemistry up to $1 \times 10^7$ years.

Figure \ref{fig:comparison_cloud} shows the ratio of the predicted abundances obtained with the cleaned \textsc{GRETOBAPE} and original reaction networks as a function of time, for the species where this ratio is larger than 3 or smaller than 0.3 at any time $\geq 10^3$ yr of the simulation and whose abundances are larger than $10^{-13}$ at steady-state. 
The figure also shows the abundances of these species, to further appreciate the impact.
Table \ref{tab:comparison_cloud_min_max} reports the minimum and maximum ratios in the time intervals $10^4$--$10^5$ yr, $10^5$--$10^6$ yr and at steady-state, respectively.

\begin{deluxetable*}{l|cc|cc|cc|cc}
\tablecaption{Results of the modeling of the cold molecular cloud (\S ~\ref{subsubsec:astro-cloud}). 
    For each species, listed in the first column, columns 2-5 list the minimum (\texttt{min}) and maximum (\texttt{max}) values of the network post- (\textsc{GRETOBAPE}) over pre-cleaning predicted abundance ratios in the time range 1$\times10^{4}$--1$\times10^{5}$ and 1$\times10^{5}$--1$\times 10^{6}$ yr, respectively.
    Column 6-7 report the ratio and the abundance at steady-state (SS).
    The last two columns list the species largest reached abundance and at what time.
    The species are ordered following the maximum ratio variation during the entire simulation.
    Finally, boldfaced species have been detected in the ISM.
}
\label{tab:comparison_cloud_min_max}
\tablehead{Interval       & \multicolumn{2}{c|}{1$\times10^{4}$--1$\times10^{5}$ yr} & \multicolumn{2}{c|}{1$\times10^{5}$--1$\times10^{6}$ yr} & \multicolumn{2}{c}{SS} & \multicolumn{2}{|c}{\texttt{max}(Abd/H)}}
\startdata
Species        & \texttt{min}(ratio)      & \texttt{max}(ratio)    & \texttt{min}(ratio)    & \texttt{max}(ratio)    &  ratio & Abd/H & Abd/H &  \texttt{time}[yr] \\
\hline
HCSi                        & 4.2$\times 10^{-2}$ & 6.2$\times 10^{-1}$ & 6.7$\times 10^{-2}$ & 8.6$\times 10^{-1}$ & 5.2$\times 10^{-2}$ & 1.1$\times 10^{-14}$ & 1.1$\times 10^{-12}$ & 7.5$\times 10^{4}$ \\
SiO$^+$                     & 3.8$\times 10^{2}$  & 1.6$\times 10^{3}$  & 1.3$\times 10^{3}$  & 2.1$\times 10^{3}$  & 2.7$\times 10^{3}$  & 8.2$\times 10^{-12}$ & 8.2$\times 10^{-12}$ & 2.4$\times 10^{6}$ \\
CHSi$^+$                    & 3.1$\times 10^{3}$  & 1.0$\times 10^{5}$  & 4.7$\times 10^{4}$  & 2.1$\times 10^{5}$  & 9.2$\times 10^{4}$  & 1.0$\times 10^{-13}$ & 2.7$\times 10^{-12}$ & 6.1$\times 10^{4}$ \\
HCCSi                       & 1.2$\times 10^{-1}$ & 5.8$\times 10^{-1}$ & 8.1$\times 10^{-3}$ & 1.1$\times 10^{-1}$ & 2.9$\times 10^{-1}$ & 5.2$\times 10^{-13}$ & 6.5$\times 10^{-12}$ & 1.3$\times 10^{5}$ \\
\hline
SiH                         & 1.2                 & 6.7                 & 4.5                 & 1.3$\times 10^1$    & 9.5                 & 2.1$\times 10^{-11}$ & 2.2$\times 10^{-11}$ & 2.6$\times 10^{3}$ \\
HNSi                        & 8.4$\times 10^{-1}$ & 2.0                 & 2.2$\times 10^{-1}$ & 3.4                 & 2.2$\times 10^{-2}$ & 1.4$\times 10^{-11}$ & 1.3$\times 10^{-10}$ & 7.1$\times 10^{5}$ \\
HSiNH$^+$                   & 8.6$\times 10^{-1}$ & 2.5                 & 1.8$\times 10^{-1}$ & 3.2                 & 1.8$\times 10^{-2}$ & 2.5$\times 10^{-14}$ & 2.9$\times 10^{-13}$ & 6.4$\times 10^{5}$ \\
\textbf{Si}$^\mathbf{+}$    & 1.4                 & 1.0$\times 10^1$    & 4.5                 & 2.4$\times 10^1$    & 9.4                 & 1.8$\times 10^{-9}$  & 7.8$\times 10^{-09}$ & 1.0$\times 10^{2}$ \\
SiF$^+$                     & 1.4                 & 1.0$\times 10^1$    & 4.3                 & 2.3$\times 10^1$    & 8.9                 & 6.6$\times 10^{-13}$ & 8.0$\times 10^{-13}$ & 9.6$\times 10^{5}$ \\
SiH$_2^+$                   & 1.4                 & 1.0$\times 10^1$    & 4.4                 & 2.3$\times 10^1$    & 9.2                 & 3.7$\times 10^{-12}$ & 4.7$\times 10^{-12}$ & 8.7$\times 10^{5}$ \\
\textbf{H$\mathbf{_2}$CS}   & 1.8$\times 10^{-2}$ & 3.7$\times 10^{-2}$ & 2.8$\times 10^{-2}$ & 6.7$\times 10^{-2}$ & 1.2$\times 10^{-1}$ & 3.4$\times 10^{-12}$ & 2.9$\times 10^{-11}$ & 1.7$\times 10^{5}$ \\
\hline
\textbf{SiC}                & 1.9                 & 5.4                 & 6.8$\times 10^{-1}$ & 5.1                 & 7.1                 & 6.0$\times 10^{-13}$ & 2.5$\times 10^{-11}$ & 2.9$\times 10^{3}$ \\
SiH$^+$                     & 1.3                 & 3.1                 & 2.0                 & 4.9                 & 7.7                 & 1.9$\times 10^{-12}$ & 2.3$\times 10^{-12}$ & 8.7$\times 10^{5}$ \\
HSiO$^+$                    & 1.5$\times 10^{-1}$ & 1.9$\times 10^{-1}$ & 1.3$\times 10^{-1}$ & 1.9$\times 10^{-1}$ & 1.5$\times 10^{-1}$ & 7.4$\times 10^{-12}$ & 7.4$\times 10^{-12}$ & 1.0$\times 10^{7}$ \\
\textbf{SiN}                & 1.3                 & 5.4                 & 1.8                 & 5.8                 & 3.1                 & 2.5$\times 10^{-12}$ & 1.8$\times 10^{-11}$ & 6.4$\times 10^{5}$ \\
\textbf{SiNC}               & 8.8$\times 10^{-1}$ & 2.8                 & 1.6                 & 5.8                 & 3.4                 & 2.5$\times 10^{-15}$ & 1.2$\times 10^{-13}$ & 6.8$\times 10^{4}$ \\
l-SiC$_3$                   & 1.1                 & 5.5                 & 3.6                 & 7.7                 & 6.5                 & 1.0$\times 10^{-14}$ & 9.0$\times 10^{-12}$ & 3.5$\times 10^{5}$ \\
SiC$_3$H                    & 5.8$\times 10^{-1}$ & 8.8$\times 10^{-1}$ & 5.8$\times 10^{-1}$ & 1.1                 & 4.5                 & 1.8$\times 10^{-15}$ & 1.5$\times 10^{-11}$ & 2.1$\times 10^{5}$ \\
SiCH$_3$                    & 8.8$\times 10^{-1}$ & 9.6$\times 10^{-1}$ & 9.9$\times 10^{-1}$ & 3.3                 & 6.4                 & 5.9$\times 10^{-14}$ & 7.6$\times 10^{-13}$ & 1.5$\times 10^{5}$ \\
l-C$_3$H$_2^+$              & 5.3$\times 10^{-1}$ & 9.0$\times 10^{-1}$ & 1.0$\times 10^{-1}$ & 4.7$\times 10^{-1}$ & 2.7$\times 10^{-1}$ & 2.5$\times 10^{-15}$ & 3.7$\times 10^{-11}$ & 1.0$\times 10^{5}$ 
\enddata
\tablecomments{The detected molecules are bold-faced \citep{McGuireCensus2021}. 
The Si$^+$ atom was first detected by \citet{haas1986detection}.}
\end{deluxetable*}

Overall, 26 species are significantly affected by the reaction network cleaning process, meaning that their abundances have changed by more than a factor three during the evolution using the two reaction networks.
The affected species represent only 5\% of the species in the clean network \textsc{GRETOBAPE}, which is a reassuring result.

The species impacted by the reaction network cleaning process can be divided into three groups, where we only consider those whose abundance is larger than $10^{-13}$ during the evolution:
\begin{itemize}
    \item [1.] The first group (upper panels of Fig.~\ref{fig:comparison_cloud}) contains species where the post- over pre- cleaning ratio is larger than $10^3$ or lower than $10^{-3}$ during the evolution: HCSi, SiO$^+$, CHSi$^+$ and HCCSi.
    None of these species have been detected in space.
    Note that five of these six species contain Si and all of them have an abundance lower than $2\times 10^{-11}$.
    The most abundant one is SiO$^+$, with a predicted steady-state abundance of $8\times 10^{-12}$, $3\times 10^3$ larger with respect to the original reaction network. 
    The species whose ratio is the largest at steady-state is CHSi$^+$, when its abundances is lower than $10^{-13}$.
    HCCSi presents a ratio of 0.3 and an abundance of $5 \times 10^{-13}$ at steady-state, while at $\sim 2 \times 10^{5}$ yr the minimum ratio value is 0.008. 
    We will discuss in \S ~\ref{subsubsec:astro-specific-cases} the Si chemistry and why it is so impacted by the reaction network cleaning process.
    \item [2.] The second group (middle panels of Fig.~\ref{fig:comparison_cloud}) contains species where the ratio is  within 10 and $10^3$ or $10^{-3}$ and 0.1 during the evolution: SiH, HNSi, HSiNH$^+$, Si$^+$, SiF$^+$, SiH$_2^+$ and H$_2$CS.
    H$_2$CS and Si$^+$ are the only species detected in space belonging to this group. 
    H$_2$CS has a ratio ranging from $0.1$ to $0.03$ during the evolution, namely its abundance is lower in the cleaned network simulation by about a factor 10 at steady-state ($3.4\times 10^{-12}$ instead of $4\times 10^{-13}$).
    The cleaning process affects identically, with the same ratio profile and increasing the abundances by one order of magnitude at steady-state, the following Si-bearing species: SiH, Si$^+$, SiF$^+$ and SiH$_2^+$. 
    HNSi and HSiNH$^+$ share the same behavior, but with a steady-state reduced abundance by about a factor 100.
    The predicted SO$_2^+$ abundance is a factor $\sim 10$ larger with the new network.
    \item [3.] The third group (bottom panels of Fig.~\ref{fig:comparison_cloud}) contains species where the ratio is within 3 and 10 or 0.1 and 0.3: SiC, SiH$^+$, HSiO$^+$ and SiN.
    Three species of this group have been detected in the ISM: SiC, SiN and SiNC. 
    At steady-state, the abundance of SiC is about a factor 10 larger with the new reaction network, while that of SiN is a factor $\sim$3 larger.
\end{itemize}

Overall, 5 out of 26 species affected by the cleaning process have been detected in the ISM: Si$^+$, SiC, SiN, SiNC and H$_2$CS.
Of those, only thioformaldehyde (H$_2$CS) has been detected so far in cold molecular clouds \citep{irvine1989observations, minowa1997laboratory, marcelino2005deuterated}.
We will discuss the case of H$_2$CS in \S~\ref{subsubsec:astro-specific-cases}.
Finally, 18 out of 20 species affected by the cleaning process contain Si.
We will discuss in some detail this case in \S~\ref{subsubsec:astro-specific-cases}.

\subsubsection{Warm molecular outflow shock}
\label{subsubsec:astro-shock}

Young protostars are known to have spectacular ejections of matter which cause shocks when they hit the surrounding quiet environment.
From the chemical point of view, the passage of the shock has two major effects: (i) it heats and compresses the gas and (ii) species frozen on the icy grain mantles are sputtered and injected into the gas phase.
As a result, molecular shocks have gas warm ($\sim$80--100 K) and dense ($\geq 10^5$ cm$^{-3}$) regions with much larger abundances of some species with respect to cold molecular clouds.

In order to simulate the chemical composition of the gas after the passage of a shock, we adopt a two-step procedure where we first compute the chemical abundances of the cold molecular cloud (previous subsection) and then we suddenly (i) increase the gas temperature and the H nuclei number density and (ii) inject into the gas large abundances of species that are known to be present on the grain icy mantles.
Although the physics is slightly different, a similar modeling can be also considered a fair enough simulation of what happens in hot cores and hot corinos, which are regions heated by the central forming star and where the grain icy mantles sublimate \citep[\textit{e.g.} see the recent review by][]{Ceccarelli2022}.
Also, we chose to assume the pre-shock abundances from a steady state cold cloud because our goal is not to reproduce a specific case, but just to approximately understand the impact of the endothermicity correction in a “warm” case. 
In this respect, the steady state choice has two advantages: it does not depend on the choice of the time (i.e. choice of initial abundances) and it only depends on the change in the reaction network.

In practice, our model has two steps, as follows.\\
\underline{Step 1:} We first compute the steady-state chemical composition of a cold
molecular cloud at 10 K with a H nuclei number density of $2\times 10^4$ cm$^{-3}$ and with the parameters described in \S~\ref{subsubsec:astro-cloud}.\\
\underline{Step 2:} We increase the gas temperature to 90 K, H nuclei number density to
$8 \times 10^5$ cm$^{-3}$ and the gaseous abundance of grain-mantle species as follows: 
CO$_2$/H = $3\times 10^{-5}$, 
H$_2$O/H = $2\times 10^{-4}$,
NH$_3$/H = $2\times 10^{-5}$,
CH$_3$OH/H = $6.5\times 10^{-6}$,
CH$_3$CHO/H = $3.8\times 10^{-8}$,
C$_2$H$_5$/H = $8\times 10^{-8}$,
SiO/H = $1\times 10^{-6}$,
Si/H = $1\times 10^{-6}$
and
OCS/H = $2\times 10^{-6}$.
We then leave the chemical composition evolve for $10^4$ yr.

The values adopted for the physical parameters and abundances of the injected species are based on the astronomical observations of the prototypical molecular outflow shock L1157-B1, for which our group carried the same model described here \cite[\textit{e.g.}][]{Codella_2017_SOLIS_II, Podio2017-SiS, Codella_2020_SOLISV}.

The results of the modeling are reported in Tab. \ref{tab:comparison_shock_min_max}, which reports the ratio of the predicted abundances obtained with the cleaned \textsc{GRETOBAPE} and original reaction networks in the time intervals $1\times 10^2$--$1\times 10^3$ yr and $1\times 10^3$--$5\times10^3$ yr for species with a ratio larger than 3 or less than 0.3 and abundances larger than $10^{-13}$. 

\begin{deluxetable*}{l|cc|cc|cc|cc}
\tablehead{Interval       & \multicolumn{2}{c|}{1$\times10^{2}$--1$\times10^{3}$ yr} & \multicolumn{2}{c|}{1$\times10^{3}$--5$\times 10^{3}$ yr} & \multicolumn{2}{c|}{5$\times 10^{3}$ yr} & \multicolumn{2}{|c}{\texttt{max}(Abd/H)}}
    \tablecaption{Results of the modeling of the warm molecular outflow shock (\S ~\ref{subsubsec:astro-shock}). 
    For each species, listed in the first column, columns 2-5 list the minimum (\texttt{min}) and maximum (\texttt{max}) values of the network post- (\textsc{GRETOBAPE}) over pre-cleaning predicted abundance ratios in the time range 1$\times10^{2}$--1$\times10^{3}$ and 1$\times10^{3}$--5$\times 10^{3}$ yr, respectively.
    Column 6-7 report the ratio and the abundance at $5\times 10^{3}$ yr.
    The last two column list the species largest reached abundance and at what time.
    The species are ordered following the value of the ratio at 5$\times10^{3}$ yr, which is believed to approximately be the L1157-B1 age (\S~\ref{subsubsec:astro-shock}).
    Finally, boldfaced species have been detected in the ISM.
}
\label{tab:comparison_shock_min_max}
\startdata
Species        & \texttt{min}(ratio)        & \texttt{max}(ratio)        &  \texttt{min}(ratio)       & \texttt{max}(ratio)        & ratio                & Abd/H              & Abd/H                &  \texttt{time}[yr] \\
\hline
SiS$^+$        & 3.9$\times 10^{7}$  & 4.2$\times 10^{7}$  & 4.2$\times 10^{7}$  & 4.5$\times 10^{7}$  & 4.2$\times 10^{7}$  & 3.9$\times 10^{-11}$ & 6.3$\times 10^{-11}$ & 1.0$\times 10^{2}$ \\
CHSi$^+$       & 1.5$\times 10^{6}$  & 1.6$\times 10^{6}$  & 1.4$\times 10^{6}$  & 1.9$\times 10^{6}$  & 1.4$\times 10^{6}$  & 4.3$\times 10^{-13}$ & 4.3$\times 10^{-13}$ & 4.9$\times 10^{3}$ \\
SiO$^+$        & 1.3$\times 10^{3}$  & 1.5$\times 10^{3}$  & 1.6$\times 10^{3}$  & 7.9$\times 10^{3}$  & 7.9$\times 10^{3}$  & 1.1$\times 10^{-11}$ & 1.1$\times 10^{-11}$ & 5$\times 10^{3}$ \\
\textbf{SiN}   & 1.2$\times 10^{-2}$ & 1.9$\times 10^{-2}$ & 2.8$\times 10^{-3}$ & 7.7$\times 10^{-3}$ & 5.5$\times 10^{-3}$ & 3.6$\times 10^{-13}$ & 3.6$\times 10^{-13}$ & 5$\times 10^{3}$ \\
HSiO$^+$       & 1.1$\times 10^{-1}$ & 1.1$\times 10^{-1}$ & 8.3$\times 10^{-2}$ & 1.0$\times 10^{-1}$ & 8.2$\times 10^{-2}$ & 3.5$\times 10^{-10}$ & 3.5$\times 10^{-10}$ & 3.9$\times 10^{3}$ \\
\hline
\textbf{CH$_3$CHCH$_2$} & 5.0$\times 10^{-2}$ & 5.2$\times 10^{-2}$ & 5.3$\times 10^{-2}$ & 6.5$\times 10^{-2}$ & 6.5$\times 10^{-2}$ & 3.6$\times 10^{-13}$ & 4.2$\times 10^{-13}$ & 1.0$\times 10^{2}$ \\
SiH$_2^+$      & 4.0$\times 10^{2}$  & 4.3$\times 10^{2}$  & 4.4$\times 10^{2}$  & 5.0$\times 10^{2}$  & 4.9$\times 10^{2}$  & 1.5$\times 10^{-12}$ & 1.5$\times 10^{-12}$ & 5$\times 10^{3}$ \\
\textbf{Si}$^+$& 4.0$\times 10^{2}$  & 4.3$\times 10^{2}$  & 4.4$\times 10^{2}$  & 4.9$\times 10^{2}$  & 4.8$\times 10^{2}$  & 1.5$\times 10^{-10}$ & 1.6$\times 10^{-10}$ & 3.9$\times 10^{3}$ \\
SiH            & 7.1$\times 10^{1}$  & 1.4$\times 10^{2}$  & 1.6$\times 10^{2}$  & 3.7$\times 10^{2}$  & 3.7$\times 10^{2}$  & 3.9$\times 10^{-12}$ & 3.1$\times 10^{-12}$ & 4.3$\times 10^{3}$ \\
HNSi           & 1.6$\times 10^{-2}$ & 1.9$\times 10^{-2}$ & 1.1$\times 10^{-2}$ & 1.5$\times 10^{-2}$ & 1.1$\times 10^{-2}$ & 1.3$\times 10^{-11}$ & 1.4$\times 10^{-11}$ & 8.0$\times 10^{2}$ \\
SiNC$^+$       & 5.5$\times 10^{1}$  & 7.7$\times 10^{1}$  & 8.4$\times 10^{1}$  & 1.2$\times 10^{2}$  & 1.1$\times 10^{2}$  & 2.5$\times 10^{-12}$ & 2.4$\times 10^{-12}$ & 4.7$\times 10^{3}$ \\
\hline
\textbf{HSS}   & 9.6$\times 10^{-1}$ & 1.0                 & 2.3$\times 10^{-1}$ & 9.2$\times 10^{-1}$ & 2.3$\times 10^{-1}$ & 4.7$\times 10^{-13}$ & 4.7$\times 10^{-13}$ & 5$\times 10^{3}$ \\
c-HCCHSi       & 4.3$\times 10^{-1}$ & 1.2                 & 2.3$\times 10^{-1}$ & 3.7$\times 10^{-1}$ & 2.3$\times 10^{-1}$ & 6.2$\times 10^{-12}$ & 6.1$\times 10^{-12}$ & 4.9$\times 10^{3}$ \\
\textbf{c-SiC$_2$}& 4.3$\times 10^{-2}$ & 5.5$\times 10^{-2}$ & 6.3$\times 10^{-2}$ & 2.1$\times 10^{-1}$ & 2.1$\times 10^{-1}$ & 7.2$\times 10^{-13}$ & 7.3$\times 10^{-13}$ & 4.7$\times 10^{3}$ \\
\textbf{SiS}   & 1.7$\times 10^{-1}$ & 2.8$\times 10^{-1}$ & 1.5$\times 10^{-1}$ & 1.7$\times 10^{-1}$ & 1.5$\times 10^{-1}$ & 8.4$\times 10^{-12}$ & 8.5$\times 10^{-12}$ & 4.9$\times 10^{3}$ \\
SO$_2^+$       & 1.3$\times 10^{1}$  & 1.4$\times 10^{1}$  & 1.4$\times 10^{1}$  & 1.5$\times 10^{1}$  & 1.5$\times 10^{1}$  & 1.4$\times 10^{-13}$ & 1.4$\times 10^{-13}$ & 5$\times 10^{3}$ \\
HCCSi          & 2.4$\times 10^{-2}$ & 3.8$\times 10^{-2}$ & 4.4$\times 10^{-2}$ & 1.3$\times 10^{-1}$ & 1.3$\times 10^{-1}$ & 4.8$\times 10^{-12}$ & 4.7$\times 10^{-12}$ & 4.9$\times 10^{3}$ \\
Si             & 1.6$\times 10^{-1}$ & 1.7$\times 10^{-1}$ & 1.2$\times 10^{-1}$ & 1.6$\times 10^{-1}$ & 1.2$\times 10^{-1}$ & 3.3$\times 10^{-10}$ & 2.1$\times 10^{-09}$ & 1.0$\times 10^{2}$ \\
HSSH           & 8.0$\times 10^{-1}$ & 1.0                 & 1.0$\times 10^{-1}$ & 6.7$\times 10^{-1}$ & 1.0$\times 10^{-1}$ & 1.6$\times 10^{-13}$ & 1.6$\times 10^{-13}$ & 5$\times 10^{3}$ \\
SiH$^+$        & 2.4                 & 3.3                 & 3.7                 & 5.7                 & 5.1                 & 5.3$\times 10^{-13}$ & 5.4$\times 10^{-13}$ & 4.9$\times 10^{3}$ \\
\textbf{SiC}   & 1.2                 & 1.2                 & 1.2                 & 3.0                 & 3.0                 & 2.3$\times 10^{-11}$ & 2.3$\times 10^{-11}$ & 5$\times 10^{3}$  \\
\textbf{C$_3$O}& 9.8                 & 2.1$\times 10^{1}$  & 2.0                 & 1.6$\times 10^{1}$  & 2.0                 & 2.4$\times 10^{-12}$ & 7.9$\times 10^{-12}$ & 1.0$\times 10^{2}$ 
\enddata
\tablecomments{The detected species are bold-faced \citep{McGuireCensus2021}. 
The Si$^+$ atom was first detected by \citet{haas1986detection}.}
\end{deluxetable*}

Overall, 22 species are significantly  affected by the reaction network cleaning process, namely their abundances change by more than a factor 3 during the evolution from the passage of the shock and $5\times10^3$ yr.
The species impacted by the reaction network cleaning process can be divided into three groups with respect to their ratio at $5\times10^3$ yr (\textit{e.g.} the approximate L1157-B1 age):
\begin{itemize}
    \item [1.] The first group contains species whose ratio is larger than $10^{3}$ or lower $10^{-3}$: SiS$^+$, CHSi$^+$, SiO$^+$ and SiN. 
    SiN is the only species in this group that has been detected in outflows and will be discuss in detail in a dedicated paragraph of \S ~\ref{subsubsec:astro-specific-cases}.
    The species with the largest variation due to the cleaning process is SiS$^+$, with an almost constant ratio of $4.2\times10^7$ and reaching an abundance of $\sim 4\times10^{-11}$ at $5\times10^3$ yr.
    \item [2.] The second group contains species where the ratio is within $10^{2}$ and $10^{3}$ or $10^{-3}$ and $10^{-2}$: HSiO$^+$, CH$_3$CHCH$_2$, SiH$_2^+$, Si$^+$, SiH, HNSi, SiNC$^+$.
    Two species of this group have been detected in the ISM: CH$_3$CHCH$_2$ and Si$^+$. 
    Only the latter has been detected in outflows and we discuss its case in \S ~\ref{subsubsec:astro-specific-cases}.
    CH$_3$CHCH$_2$ has an almost constant ratio of $\sim 5\times10^{-2}$ and a maximum abundance of $\sim 5\times10^{-13}$ at the beginning of the simulation. 
    On the other hand Si$^+$ has a ratio of almost $4.5\times10^{2}$ that increases the abundance in the cleaned simulation with respect to the original.
    \item [3.] The third group contains species where the ratio is within 3 and $10^{2}$ or $10^{-2}$ and 0.3 during the evolution: HSS, c-HCCHSi, c-SiC$_2$, SiS, SO$_2^+$, HCCSi, Si, HSSH, SiH$^+$, SiC and C$_3$O. 
    Five species belonging to this group have been detected in the ISM: HSS, c-SiC$_2$, SiS, SiC and C$_3$O.
    Only SiS has been detected in outflows and we discuss its case in \S ~\ref{subsubsec:astro-specific-cases}.
    HSS has an abundance ratio between 0.2 and 0.9 in the $1\times10^3$--$5\times10^3$ yr interval and a maximum abundance of 4.7$\times10^{-13}$ at the end of the simulation.
    c-SiC$_2$ at the L1157-B1 age has a ratio of 2$\times10^{-1}$ and an abundance of 7.2$\times10^{-13}$.
    SiS has an almost constant ratio of 0.15 meanwhile C$_3$O ratio oscillates between $\sim 20$ and 2.
    SiC is the less affected species by the cleaning process with a factor 3 of difference.
\end{itemize}

As in the cold molecular cloud simulation (\S~\ref{subsubsec:astro-cloud}), the cleaning process mostly affected the Si reaction network.
Indeed, 17 of the 22 species present in Tab. \ref{tab:comparison_shock_min_max} are Si-bearing species. 
Four of them, SiN \citep{schilke2003interstellar}, SiS \citep{Podio2017-SiS} and Si$^+$ \citep{haas1986detection} are detected in outflows.
They will be discussed in some detail in \S~\ref{subsubsec:astro-specific-cases}.

\subsubsection{Discussion of specific cases}\label{subsubsec:astro-specific-cases}

In this section, we discuss in some details the species that are more significantly affected by the reaction network cleaning process.

\paragraph{Thioformaldehyde}

In cold molecular gas (\S ~\ref{subsubsec:astro-cloud}), the thioformaldehyde (H$_2$CS) abundance predicted after the cleaning process at the steady-state is about 10 times smaller than in pre-cleaning.
Here we first summarize the observations and then we comment on what causes the difference in the predicted abundances.

Thioformaldehyde has been detected in cold molecular clouds, notably TMC-1 and L134, with an abundance (wrt to H nuclei) of about $\sim 10^{-9}$ \citep[\textit{e.g.}][]{Ohishi1998}.
\cite{vastel2018} and \cite{Spezzano2022} carried out a detailed study of thioformaldehyde toward the prototypical cold prestellar core L1544, reporting a column density of $\sim 6\times 10^{12}$ cm$^{-2}$, corresponding to an average abundance of $\leq 10^{-10}$, depending on where the detected H$_2$CS emission comes from.
These authors attribute the relatively low abundance of all the S-bearing species observed in these objects to a general depletion of the gaseous S elemental abundance.
Likewise, \cite{Esplugues2022} have carried out a detailed study of H$_2$CS towards several cold starless cores in the Taurus, Perseus, and Orion regions and found H$_2$CS abundances ranging from 0.8 to 14 $\times 10^{-11}$.

As discussed in detail in \cite{Esplugues2022}, in cold gas, H$_2$CS is mainly formed by two reactions:
S + CH$_3$ and H$_3$CS$^+$ + e$^-$.
While H$_2$CS is not directly involved in the cleaning process (Tab. \ref{tab:del_mol_info}), the H$_3$CS$^+$ post-cleaning abundance at steady-state is lower by a factor $\sim 18$ than the pre-cleaned one because one dominant reaction in the original network forming H$_3$CS$^+$, CH$4$ + S$^+$ $\rightarrow$ H + H$_3$CS$^+$, is endothermic by 48 kJ/mol.
Please note that the steady-state post-cleaning abundance of H$_3$CS$^+$ is $\leq10^{-13}$ so that this species does not appear in Fig. \ref{fig:comparison_cloud} and Tab. \ref{tab:comparison_cloud_min_max}.
However, the low abundance ($6\times 10^{-15}$) is counterbalanced by the large recombination rate of H$_3$CS$^+$, whose product is H$_2$CS.

\paragraph{Si chemistry and Si$^+$}
As already anticipated, the reaction network cleaning process significantly affects the Si chemistry, with the removal of 48 reactions and 6 species that represent the $\sim 9$\% and $\sim 10$\% of total Si-bearing reactions and species, respectively.
In addition, 18 out of 20 most affected species of the cold molecular cloud simulation (\S ~\ref{subsubsec:astro-cloud}) contain Si, of which the vast majority (12) have less than four atoms. 
Similarly, 17 of the 22 most affected species in the the warm molecular outflow shock simulation (\S ~\ref{subsubsec:astro-shock}) are Si-bearing species.

The six removed Si-bearing species (Tab. \ref{tab:del_mol_info}) are (neutral) chains and are removed because their formation or destruction routes in the original network are endothermic (\S ~\ref{subsubsec:cleaning-sources-errors}).
Their removal very weakly affects two other Si-bearing species,  c-HCCHSi and SC$_3$H, respectively (Figs.~\ref{fig:Si_network} and \ref{fig:Si_deleted_network}) and, consequently, the rest of the Si-bearing species.

Therefore, the large impact of the cleaning process on the Si chemical network is due to the removal of the 48 reactions involving Si-bearing species.
Figure \ref{fig:Si_deleted_network} shows that those reactions connect 36 Si-bearing species out of 61, namely more than half of them.
It is, therefore, not surprising that 17 Si-bearing species are among the most affected ones in the warm molecular outflow shock model.
For example, SiS$^+$, the most affected species, has three removed reactions connecting it to Si$^+$, SiC$_3$H$_5$ and SiS, respectively.
The first two are destruction routes of SiS$^+$; their removal is, therefore, responsible of the augmented predicted SiS$^+$ abundance.
A similar argument applies to CHSi$^+$, the second most affected species in the modeling: a reaction of destruction is removed, causing the increase in the predicted abundance.

Probably most important, Si$^+$ has 10 removed reactions linking it to other Si-bearing species.
Because of that, the predicted abundance of Si$^+$ is larger more than 400 times the predicted abundance before the cleaning process.
This causes a domino effect, propagating as a wave: 11 of the other 16 impacted Si-bearing species are one step away from Si$^+$ (\textit{i.e.} 1 node distance in Fig. \ref{fig:Si_network}), while the other 5 are two steps away.
The species one step away are also those whose predicted abundances are the most impacted.

\paragraph{Silicon nitride} 
Silicon nitride, SiN, is the detected Si-bearing species most affected by the cleaning process (Tab.~\ref{tab:comparison_shock_min_max}), so we will analyze its case in some detail here.
SiN was first detected in the circumstellar envelope of the evolved star IRC+10216 \citep{turner1992detection} and, later towards the Galactic Center cloud SgrB2(M) \citep{schilke2003interstellar}, probably in the warm shocked gas of the region. 
IRC+10216 and SgrB2(M) remain the only two objects where SiN has so far been detected, despite sensitive spectral surveys towards the high-mass star forming region Orion \citep{Tercero2011-SiN} and several low-mass star forming regions, including the prototype of the warm molecular shocks L1157-B1 \citep{Ceccarelli2017-solis}.

The predicted SiN abundance is more than 100 times lower when using the post-cleaning reaction network than the pre-cleaning one.
Carefully looking at Figs. \ref{fig:Si_network} and \ref{fig:Si_deleted_network}, one can notice that the SiN major formation route in the pre-cleaning network is the reaction involving HSiNH$^+$, followed by those involving SiC and HNSi$^+$ at a lesser extent.
Although this reaction is not affected by the cleaning process, two reactions forming HSiNH$^+$ are removed after it (from Si$^+$ and HNSi, respectively).  
Indeed, the abundance of HSiNH$^+$ changes by three orders of magnitude (but it is not listed in Tab. \ref{tab:comparison_shock_min_max} because of its low abundance, $\leq 10^{-13}$). 
About these reactions, as discussed for the HSiO$^+$ case (see \S~\ref{par:missing_destruction}), it is probable that KIDA erroneously considered the product HSiNH$^+$ instead of its more stable isomer SiNH$_2^+$ \citep{parisel1996interstellar,glosik1995selected}. 

As said, the new reaction network predicts a much lower SiN abundance than before, which may explain the paucity of the SiN detections.
For example, \citet{Tercero2011-SiN} found a lower limit of [SiO]/[SiN]$\geq 121$.
In our post-cleaning modeling, the SiO abundance at $5\times10^3$ yr is $\sim 2\times 10^{-6}$, implying a predicted [SiO]/[SiN]$\sim5\times 10^{-6}$, therefore largely compatible with the non detection of \cite{Tercero2011-SiN}.

\paragraph{SiS}
Silicon monosulfide, SiS, is another detected Si-bearing molecules affected by the cleaning process.
Specifically, SiS has been so far detected towards two warm molecular shocks, Orion KL \citep[][]{Tercero2011-SiN} and L1157-B1 \citep[][]{Podio2017-SiS}.
The factor between the pre- and post- cleaning predicted abundances is $\sim0.15$, again probably explaining the paucity of SiS detections.
Peering into Fig. \ref{fig:Si_network} and \ref{fig:Si_deleted_network}, in the pre- and post- cleaning networks, SiS is mainly formed from HSiS$^+$, whose abundance ($\leq 10^{-13}$, so the species not present in Tab. \ref{tab:comparison_shock_min_max}) is affected by the cleaning process, with a decrease factor of $\sim 0.06$.

\subsection{Predicted abundances with the reaction network \textsc{GRETOBAPE-red} versus \textsc{GRETOBAPE}} \label{subsec:astro-reduced-network}

To explore the reliability of our reduced network, \textsc{GRETOBAPE-red}, we performed an exhaustive parameter-space exploration and compared the steady-state abundances of all species with those obtained using the complete network, \textsc{GRETOBAPE}. 

To this end, we initialized different combinations of parameters that can affect the resulting species abundances in the cold and warm conditions present during the star-formation process. 
Namely, the varied parameters are: 
the CR ionization rate for H$_2$, with values of $3\times10^{-16}$, $3\times10^{-17}$, and $3\times10^{-18}$ s$^{-1}$;
the gas temperature, with values of 10, 50, 100, and 500 K; 
and the H nuclei number density, with values of $10^{2}$, $10^{4}$, $10^{6}$, and $10^{8}$ cm$^{-3}$. 
All possible combinations of these parameters yield 48 different runs, the results of which we will explain in what follows.
We used the \textsc{krome} chemistry package \citep{Grassi2014} to perform the parameter space exploration, considering only gas-phase chemistry.
We keep the rest of the input parameters (visual extinction, dust-to-gas mass ratio, grain radius and grain density), equal to those used to perform the run described in \S~\ref{subsubsec:astro-cloud}, and left the chemistry evolve up to $1\times10^7$ years.

In all runs, the reduced network \textsc{GRETOBAPE-red} reproduces almost perfectly the steady-state abundances of \textsc{GRETOBAPE}, for all species with abundances $>10^{-10}$, with differences smaller than a factor 3 in general. 
However, the differences progressively worsen for trace species. 
A few species with abundances $\leq10^{-10}$ have differences larger than a factor 3 when comparing the results of \textsc{GRETOBAPE-red} with those of \textsc{GRETOBAPE}. 
The main species affected are the carbon chains C$_6$, C$_3$, H$_2$CCCO, c-C$_3$H$_2$O, C$_3$O, H$_2$CCN, H$_2$CCO, in addition to CH$_2$NH and O$_2$H. 
Larger differences might be present for other trace species with abundances $<10^{-15}$ but, as these abundances are already very low, these molecules may not be of interest in the astrochemical simulations where the reduced network \textsc{GRETOBAPE-red} is suitable. 
As expected, the effect on carbon chains can be attributed to the removal of larger species that can either be part of their destruction or formation routes in the case of dissociative recombination of large cations that could produce the species listed here.

These results overall indicate that the reduced network \textsc{GRETOBAPE-red} is reliable to be used when following the chemistry of species with fractional abundances larger than $\sim10^{-10}$, while the complete network \textsc{GRETOBAPE} is needed when exploring the chemistry of trace species.

\section{Conclusions} \label{sec:conclusions}

Gas-phase reaction networks are a crucial element of any astrochemical modeling.
Present day publicly available networks, \textit{e.g.} KIDA and UMIST, are made up of more than 7000 reactions, of which only a tiny fraction (10--20 \%) has been studied in laboratory or theoretical works.

In this work, we present a new gas-phase reaction network, \textsc{GRETOBAPE}, built from the publicly available KIDA2014 \citep[][]{wakelam2014_Kida_uva} with the addition of several new reactions from more recent studies.
The most important novelty is that \textsc{GRETOBAPE} is cleaned by the most obvious source of error, the presence of endothermic reactions not recognized as such.
We also present a reduced network, \textsc{GRETOBAPE-red}, to be used in problems that do not require following the chemical evolution of trace species.

To this end, we performed an extended and systematic theoretical characterization of more than 500 species at CCSD(T)/aug-cc-pVTZ//M06-2X/cc-pVTZ computational level, providing for each of them: electronic state, electronic spin multiplicity, geometry, harmonic frequency, absolute electronic energy and dipole moment. 

We then computed the enthalpy at 0 K of each neutral-neutral, neutral-ion and cation-anion reaction in the network.
Finally, we identified the endothermic reactions not recognized as such, \textit{i.e.} originally reported as barrierless or having an exponential factor $\gamma$ not taking into account the level of endothermicity. 

Following these two criteria, we deleted about 5\% of the studied reactions in the original pre-cleaned network leading to the removal of 11 species.
The final cleaned network \textsc{GRETOBAPE} consists of 6911 reactions and 488 species, while the reduced network \textsc{GRETOBAPE-red} contains 2810 reaction and 204 species.

We also reported a list of probably missing reactions, namely: 
(\textit{i}) neutral species lacking reactions with the most abundant interstellar cations (H$_3^+$, HCO$^+$, H$_3$O$^+$, He$^+$ and H$^+$), and (\textit{ii}) detected species absent in the network.

Using astrochemical model simulations of a typical cold molecular cloud and a warm molecular outflow shock, respectively, we measured the impact of the new cleaned \textsc{GRETOBAPE} on the predicted abundances.
Overall, only about 5\% of the species abundances are affected by more than a factor 3 with respect to the original pre-cleaned network predictions.
Thioformaldehyde (H$_2$CS), Si$^+$, SiN and SiS are the detected species whose predicted abundances are affected by the cleaning process.
Of the 5\% affected abundances, the immense majority concerns Si-bearing species.
We discuss in detail the origin of this large impact on the Si chemistry and conclude that a systematic review of this chemistry is needed.

We also verified that the abundances predicted by \textsc{GRETOBAPE-red} of the species with abundances larger than $10^{-10}$ are not affected by the reduction of the \textsc{GRETOBAPE} network, on a large parameter space.
Therefore, \textsc{GRETOBAPE-red} can be used when the goal is to reliably reproduce only the most abundant species and make the simulations less computational demanding.

In conclusion, we carried out the first ever systematic study of the exo/endothermicity of the reactions included in an astrochemical network.
The process removing the endothermic reactions not recognized as such and the sink/source species led to the new network \textsc{GRETOBAPE} and its reduced version \textsc{GRETOBAPE-red}.
The control of the exo/endothermicity and sink/source species is only the first and obvious step to make the astrochemical gas-phase reaction networks reliable.
Inclusion of the reactions of neutrals with abundant interstellar cations is a second obvious but still not complete step.
The next step would be to carefully review all the reactions, rate constants and products, which is obviously unfeasible given the too large number of reactions present in the networks.
A compromise could be to start with the control/verification of sub-networks that form commonly detected species, like the work done by \citet{Vazart_Cecarrelli_2020_acetaldehyde} on acetaldehyde.
In addition, the lack of gas-phase reactions forming interstellar detected species does not necessarily mean that they do not exist and a lot of work here is still necessary \citep[\textit{e.g.}][just to mention some recent works]{Balucani_2015_gasnet,Skouteris_2018_Ethanol,Sleiman2018-cyanamide, Cernicharo2022-fulvenallene}.
Last, but not least, in addition to devote studies on the formation routes, attention should also be dedicated to the destruction routes.
For example, the study of the destruction of methyl formate and dimethyl ether by He$^+$ has shown that the rate constants included in the KIDA and UMIST networks are wrong by at least a factor 10 \citep{Ascenzi_2019_dimethyl}.

Finally, the database created in this work, and available on-line (\S ~\ref{sec:online-db}), can be used to rapidly test or search existing or new possible formation and destruction reaction pathways and further refined via experiments or theoretical studies to computed rate constants.

\section{Online Database}\label{sec:online-db}
All scripts and data used in this work are publicly available in the following GitHub link: \url{https://github.com/TinacciL/GreToBaPe_Cleaning}. 
The networks and the database of all molecular species properties are available on the \href{https://aco-itn.oapd.inaf.it/aco-public-datasets/theoretical-chemistry-calculations/cations-database}{ACO (AstroChemical Origins) Database}.

\section{Supporting Material}\label{sec:support_material}

All the reaction networks used and produced in this work, and cited in the main body, are reported in the Supporting Material. 
We also provide all the \texttt{.XYZ} files of each studied species as well as all the relative information in the \texttt{Database\_Molecules.csv} file, following \cite{ISM_cations}.
Table \ref{tab:support_material} summarizes the list of files with the various reaction networks along with a brief description of their content.

\begin{deluxetable*}{ll}
\tablecaption{Summary of the files containing the reaction networks and species chemical data provided in the Supporting Material and a brief description of their content.}
\label{tab:support_material}
\tablehead{File Name & Description}
\startdata
\texttt{ReadMe.txt}                              & File with the full description of all the files in the Supporting Material.\\
\texttt{\textsc{GRETOBAPE-pre}.dat}              & Original reaction network, before the cleaning process, described in \S~\ref{subsec:results-original-newtork}. \\
\texttt{\textsc{GRETOBAPE}.dat}                  & New reaction network, after  the cleaning process, described in \S~\ref{subsec:results-new-newtork}. \\
\texttt{\textsc{GRETOBAPE-red}.dat}              & Reduced network from \texttt{\textsc{GRETOBAPE}.dat}, described in \S~\ref{subsec:results-red-network}. \\
\texttt{\textsc{GRETOBAPE-endo}.dat}             & List of reactions deleted because of the endothermicity criteria, described in \S~\ref{subsubsec:cleaning-method_endo}. \\
\texttt{\textsc{GRETOBAPE-endo-0-10kJmol}.dat}   & List of reactions with an endothermicity lower than 10 kJ/mol: they are present in GRETOBAPE.dat. \\
\texttt{\textsc{GRETOBAPE-domino}.dat}      & List of reactions deleted because of the domino effect, described in \S~\ref{subsubsec:cleaning-domino}. \\
\texttt{\textsc{GRETOBAPE-domino-endo-Si}.dat}   & List of reactions involving Si-bearing species and deleted because of the endothermicity and domino\\
                                                 & effect criteria, discussed in \S~\ref{subsubsec:cleaning-sources-errors}.\\
\texttt{Database\_Molecules.csv}                 & Extracted and organized chemical information of all species in \texttt{\textsc{GRETOBAPE}.dat}, described in \ref{subsec:meth-QM-calc}.\\
\texttt{species.zip}                             & Zip file containing the .XYZ files of all species studied in this work, described in \ref{subsec:meth-QM-calc}.\\
\enddata
\tablecomments{The file structure of all the reaction networks follows the KIDA format.}
\end{deluxetable*}

\acknowledgments

This project has received funding within the European Union’s Horizon 2020 research and innovation programme from the European Research Council (ERC) for the project ``The Dawn of Organic Chemistry'' (DOC), grant agreement No 741002, and from the Marie Sk{\l}odowska-Curie for the project ``Astro-Chemical Origins'' (ACO), grant agreement No 811312. 
SP, NB, PU acknowledge the Italian Space Agency for co-funding the Life in Space Project (ASI N. 2019-3-U.O).
CINES-OCCIGEN HPC is kindly acknowledged for the generous allowance of super-computing time through the A0060810797 project. 
We warmly thank an anonymous referee whose valuable and expert comments and suggestions helped to clarify the original manuscript.
LT is grateful to Jacopo Lupi and Stefano Ferrero for insightful discussions and to the \LaTeX $\,$ community for the insights on TikZ and PGFPlots packages. 
Finally, we wish to acknowledge the extremely useful discussions with Prof. Gretobape.

\software{\textsc{NetworkX} \citep{SciPyProceedings_11}, Gaussian16 \citep{g16}, \textsc{krome} \citep{Grassi2014}.}

\newpage

\bibliography{Tinacci-cleaning-ApJS}{}
\bibliographystyle{aasjournal}

\end{document}